\documentclass{aa}

\usepackage{graphicx}
\usepackage{txfonts}
\usepackage{amsmath}
\usepackage{hyperref}

\hypersetup{
    colorlinks = true,
    linkcolor = {blue}, 
    citecolor = {blue},
    urlcolor = {blue}
}

\usepackage{color}
\usepackage{float}
\usepackage{xspace}

\newcommand{\spitzer}{\textsl{Spitzer}\xspace}

\newcommand{\kepler}{\textsl{Kepler}\xspace}
\newcommand{\lmax}{\ensuremath{\ell_\mathrm{max}}}
\newcommand{\hml}{\ensuremath{h_{m\ell}}\xspace}
\newcommand{\ylm}{\ensuremath{Y_{\ell}^{m}}\xspace}

\begin{document}

\title{Physically-motivated basis functions for temperature maps of exoplanets}
\author{Brett M.~Morris\inst{1}
\and
Kevin Heng\inst{1,2,3}
\and
Kathryn Jones\inst{1,6}
\and
Caroline Piaulet\inst{4}
\and
Brice-Olivier Demory\inst{1}
\and
Daniel Kitzmann\inst{1}
\and
H.~Jens Hoeijmakers\inst{5}
}

\institute{
Center for Space and Habitability, University of Bern, Gesellschaftsstrasse 6, 3012 Bern, Switzerland
\and
Department of Physics, Astronomy \& Astrophysics Group, University of Warwick, Coventry CV4 7AL, United Kingdom
\and
Ludwig Maximilian University, University Observatory Munich, Scheinerstrasse 1, Munich D-81679, Germany
\and
Institut de recherche sur les exoplan\`{e}tes, D\'{e}partement de physique, Universit\'{e} de Montr\'{e}al,
2900 boul. \'{E}douard-Montpetit, Montreal QC H3T 1J4, Canada
\and
Lund Observatory, Department of Astronomy and Theoretical Physics, Lunds Universitet, Solvegatan 9, 222 24 Lund, Sweden
\and
Hans Sigrist Foundation PhD Fellow}

   \date{Received 2021; accepted 2021}

  \abstract
  {Thermal phase curves of exoplanet atmospheres have revealed temperature maps as a function of planetary longitude, often by sinusoidal decomposition of the phase curve. We construct a framework for describing two-dimensional temperature maps of exoplanets with mathematical basis functions derived for a fluid layer on a rotating, heated sphere with drag/friction, which are generalisations of spherical harmonics. These basis functions naturally produce physically-motivated temperature maps for exoplanets with few free parameters. We investigate best practices for applying this framework to temperature maps of hot Jupiters by splitting the problem into two parts: (1) we constrain the temperature map as a function of latitude by tuning the basis functions to reproduce general circulation model (GCM) outputs, since disk-integrated phase curve observations do not constrain this dimension; and (2) we infer the temperature maps of real hot Jupiters using original reductions of several \spitzer phase curves, which directly constrain the temperature variations with longitude. The resulting phase curves can be described with only three free parameters per bandpass -- an efficiency improvement over the usual five or so used to describe sinusoidal decompositions of phase curves. Upon obtaining the hemispherically averaged dayside and nightside temperatures, the standard approach would be to use zero-dimensional box models to infer the Bond albedo and redistribution efficiency. We elucidate the limitation of these box models by demonstrating that negative Bond albedos may be obtained due to a choice of boundary condition on the nightside temperature. We propose generalized definitions for the Bond albedo and heat redistribution efficiency for use with two-dimensional (2D) temperature maps. Open-source software called \textsf{kelp} is provided to efficiently compute the 2D temperature maps, phase curves, albedos and redistribution efficiencies.}
  \keywords{Infrared: planetary systems; Planets and satellites: atmospheres, gaseous planets; Techniques: photometric; Methods: analytical, observational}

   \maketitle

\section{Introduction}

We can directly observe the disk-integrated light from exoplanet atmospheres by measuring their phase curves, or the apparent brightness of the star-planet system as a function of orbital phase \citep[see review by][]{Parmentier2018}. Phase curve observations of close-in, hot exoplanets can reveal hints of atmospheric heat redistribution \citep{Knutson2007}, typically observed as a hot spot offset in longitude from the synchronously rotating sub-stellar point \citep{Cowan2018,Madhusudhan2019}. By searching for variations in the phase curve with time, one can search for signatures of non-steady-state dynamics in exoplanet atmospheres \citep[e.g.:][]{Heng2015}.

One technique for fitting the phase curves of hot exoplanets was presented by \citet{Cowan2008}, which essentially applies Fourier's theorem to phase curves. It demonstrates that a mixture of a few sinusoidal components can fit most phase curves with a relatively small number of free parameters. \citet{Cowan2008} study the information loss as one takes disk-integrated observations of exoplanet atmospheres, showing that the light curve can only sufficiently constrain roughly five free parameters. \citet{Knutson2007, Hu2015} fit the more demanding light curves of HD 189733 b and Kepler-7 b with a ``beach-ball'' model, with longitudinal segments at different temperatures. 

Several works have used spherical harmonics to describe the maps of hot exoplanets. For example, \citet{Cowan2013} derived phase curves for spherical harmonic basis maps, \citet{Rauscher2018} fit for the temperature map of HD 189733 b with a spherical harmonic expansion and a dimensionality reduction technique, and \citet{Luger2019} derive closed-form solutions for phase curves and occultations of planetary flux maps represented with spherical harmonics. One strength of these approaches is that {\it any} planetary temperature map can be represented at sufficiently large degrees, and one possible weakness is that the flexible fits may correspond to implausible climate scenarios.

\citet{Matsuno1966} and \citet{Longuet-Higgins1968} previously showed that a reduced set of two-dimensional fluid equations (the so-called ``shallow water system" or Laplace's tidal equations) on a rotating sphere may be reduced to the quantum harmonic oscillator equation. The shallow water framework has been applied to the atmospheres of neutron stars \citep{Heng2009}, the solar tachocline \citep{Gilman2000}, the atmosphere of the Earth \citep{Gill1980}, the top layer of the core of the Earth \citep{Braginsky1998} and the behaviour of submerged islands in the ocean of the Earth \citep{Longuet-Higgins1967}. In the absence of rotation, the solutions to the shallow water framework are the familiar spherical harmonics. With rotation, the solutions are the so-called ``parabolic cylinder functions", which are combinations of exponential functions and Hermite polynomials. \cite{Heng2014a} further showed that when stellar heating (forcing) and drag/friction (both hydrodynamic and magnetohydrodynamic) are added to the system, one still obtains the parabolic cylinder functions in certain limits. For brevity, we will refer to these functions as the \hml\ {\it generalised spherical harmonic basis} and use them to describe exoplanet photospheric temperature maps.

There are several advantages to using the \hml basis functions to describe an exoplanet's 2D temperature map: (1) the input arguments to the special functions, $\alpha$ and $\omega_\mathrm{drag}$, are physically-meaningful parameters which represent properties of the atmosphere; (2) it is equally straightforward to fit the 2D basis function description of the temperature map to GCM simulations and to observations of real exoplanets, allowing theory and observation to be compared in the same mathematical language; (3) even when limited to only a few free parameters, the \hml basis can reproduce a diverse range of phase curve phenomena including offset hotspots and chevron-shaped hot regions. We will show in this work that some of the physically-meaningful parameters can be constrained using fits of the \hml basis to GCMs, which reduces the total number of free parameters in fits to phase curve observations. 

In Section~\ref{sec:model} we outline the \hml basis functions used to compute the temperature map and the phase curves from the temperature map. In Section~\ref{sec:albedos} we develop formalism for measuring Bond albedos and redistribution efficiencies from 2D temperature maps. In Section~\ref{sec:gcms} we outline a technique for constraining the latitudinal temperature distributions of hot exoplanets by fitting the \hml basis functions to the 2D temperature maps produced by GCMs. In Section~\ref{sec:spitzer} we fit phase curve observations from \spitzer using priors extracted from the fits to the GCM temperature maps, and compare results obtained from fitting the \spitzer phase curves and the GCMs. We discuss the results in Section~\ref{sec:discussion} and conclude in Section~\ref{sec:conclusion}.

\section{Phase curve model} \label{sec:model}

Mathematically, \hml is the perturbation to the height of the fluid in the so-called ``shallow water system", which is a proxy for the temperature perturbation \citep{Cho2008}. Therefore, we define a 2D temperature map as a function of latitude and longitude $T(\theta, \phi)$:
\begin{equation}
    T(\theta, \phi) = \bar{T} \left( 1 + \sum_{m, \ell}^{\lmax} h_{m\ell}(\theta, \phi) \right). \label{eqn:one}
\end{equation}
The prefactor $\bar{T}$ on the right hand side acts as a constant scaling term for the mean temperature field, on which the \hml terms are a perturbation.  A plausible initial guess for the value of this prefactor is $\bar{T} = T_\star \sqrt{R_\star/a} f^\prime (1 - A_B^\prime)^{1/4}$ for stellar effective temperature $T_\star$, and normalized semimajor axis $a/R_\star$. $A_B^\prime$ is the Bond albedo and $f^\prime$ may be interpreted either geometrically (see Section \ref{sect:background}) or as a greenhouse factor. We mark the $f^\prime$ and $A_B^\prime$ variables with primes here to denote that these two terms will be redefined in Section~\ref{sec:albedos}, and thus we do not report the values of the primed variables in this work.  We note that the two terms in the factor $f^\prime (1-A_B^\prime)^{1/4}$ are perfectly degenerate. 

For efficient sampling of all plausible temperature maps given the degenerate parameterization of the two-parameter term $f^\prime(1 - A_B^\prime)^{1/4}$ in Equation~\ref{eqn:one}, we take the following conventions in generating temperature maps: (1) we set $A_B^\prime$ in Equation~\ref{eqn:one} to zero always, and (2) we allow $f^\prime$ to vary uniformly from zero to unity, which extends beyond the range normally given to this parameter in the literature. The net effects of these choices are that (1) the two free parameters $f^\prime$ and $A_B^\prime$ are sampled as a single free parameter which describes $f^\prime(1-A_B^\prime)^{1/4}$, and (2) the parameters provided to the sampler ($f^\prime\in[0, 1]$, $A_B^\prime = 0$) need not be tracked during sampling. Instead, we infer the Bond albedo and the heat redistribution efficiency for a given temperature map by computing the temperature map with the parameter $f^\prime(1-A_B^\prime)^{1/4}$, and then integrate outgoing flux from the planetary atmosphere to compute the Bond albedo and redistribution efficiency, as outlined in the Section \ref{sec:general-A_B}.

The $h_{ml}(\alpha, \omega_\mathrm{drag})$ terms are defined by equation (258) of \cite{Heng2014a},
\begin{equation}
    \begin{split}
    h_{m\ell} = \frac{C_{m\ell}}{\omega_\mathrm{drag}^2 \alpha^4 + m^2} e^{-\tilde{\mu}^2/2} [ \mu m H_{\ell} \cos(m \phi) \\
    + \alpha \omega_\mathrm{drag} (2\ell H_{\ell-1} - \tilde{\mu}H_\ell) \sin(m\phi) ],
    \end{split} \label{eqn:hml}
\end{equation}
where $\alpha$ is a dimensionless fluid number that is constructed from the Rossby, Reynolds and Prandtl numbers. $\omega_\mathrm{drag}$ is the dimensionless drag frequency (normalised by twice the angular velocity of rotation), $\mu = \cos\theta$, $\tilde{\mu}=\alpha \mu$, and $H_\ell(\tilde{\mu})$ are the physicist's Hermite polynomials: 
\begin{eqnarray}
H_0 &=& 1, \nonumber\\
H_1 &=& 2\tilde{\mu}, \nonumber\\
H_3 &=& 8\tilde{\mu}^3 - 12 \tilde{\mu}, \nonumber\\
H_4 &=& 16\tilde{\mu}^4 - 48\tilde{\mu}^2 + 12. \nonumber
\end{eqnarray}

The \hml basis is derived without any special reference point in longitude. As a result, the hottest point on the atmosphere need not be the sub-stellar point. We parameterize shifts in the prime meridian of the \hml basis from the sub-stellar point using a simple phase offset parameter, $\Delta\phi$, which describes a rotation of the coordinate system about the pole. It is important to note that the phase offset parameter $\Delta\phi$ is {\it not} equivalent to the traditional ``hotspot offset'' parameter often used in the literature, which measures the difference in longitude between the hottest longitude and the sub-stellar longitude. The phase offset $\Delta\phi$ will vary with $\omega_\mathrm{drag}$, though in practice, drag values approaching and above of $\omega_\mathrm{drag} \gtrsim 3$ correspond to maps where the hottest point on the map is practically indistinguishable from the substellar point. In the remainder of this paper, we will report the phase offset $\Delta\phi$ parameter with the implicit caveat that the distinction between phase offset and hotspot offset is negligible for the large drag frequencies discussed here.

For a phase curve fit with $\lmax=1$, for example, the free parameters in the fit could include $\{C_{11}, \alpha, \omega_\mathrm{drag}, f^\prime, \Delta \phi\}$. In practice however, $\alpha$ and $\omega_\mathrm{drag}$ control the latitudinal concentration of heat in the planet's atmosphere, so these two parameters are unconstrained by the phase curve alone. In Section~\ref{sec:gcms}, we will determine reasonable values for $\alpha$ and $\omega_\mathrm{drag}$ based on fits with the \hml basis to the 2D temperature maps produced by GCMs. These constraints can act as physically-motivated priors in the infrared phase curve fits which follow in Section~\ref{sec:spitzer}.

\begin{figure*}
    \centering
    \includegraphics[width=\columnwidth]{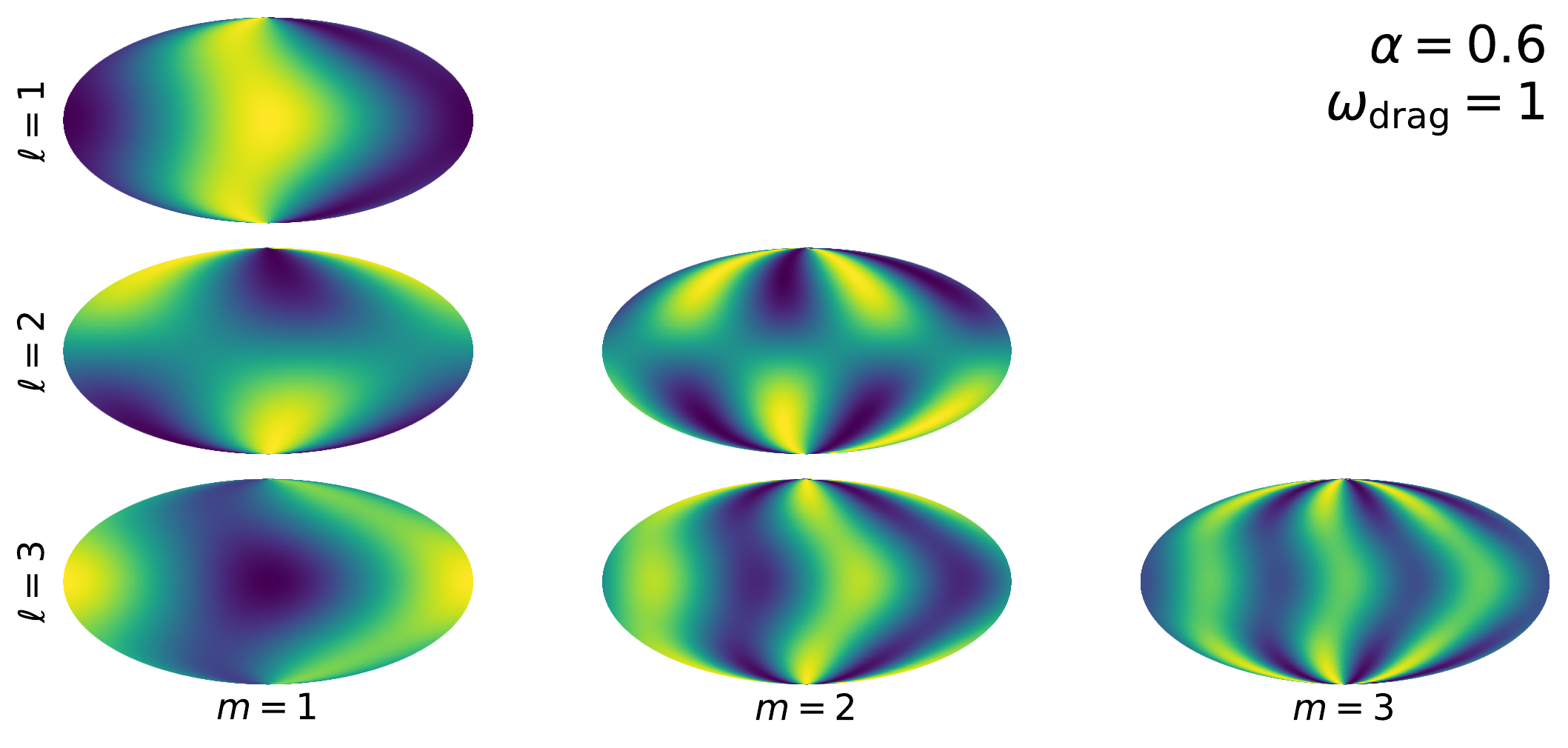}    \includegraphics[width=\columnwidth]{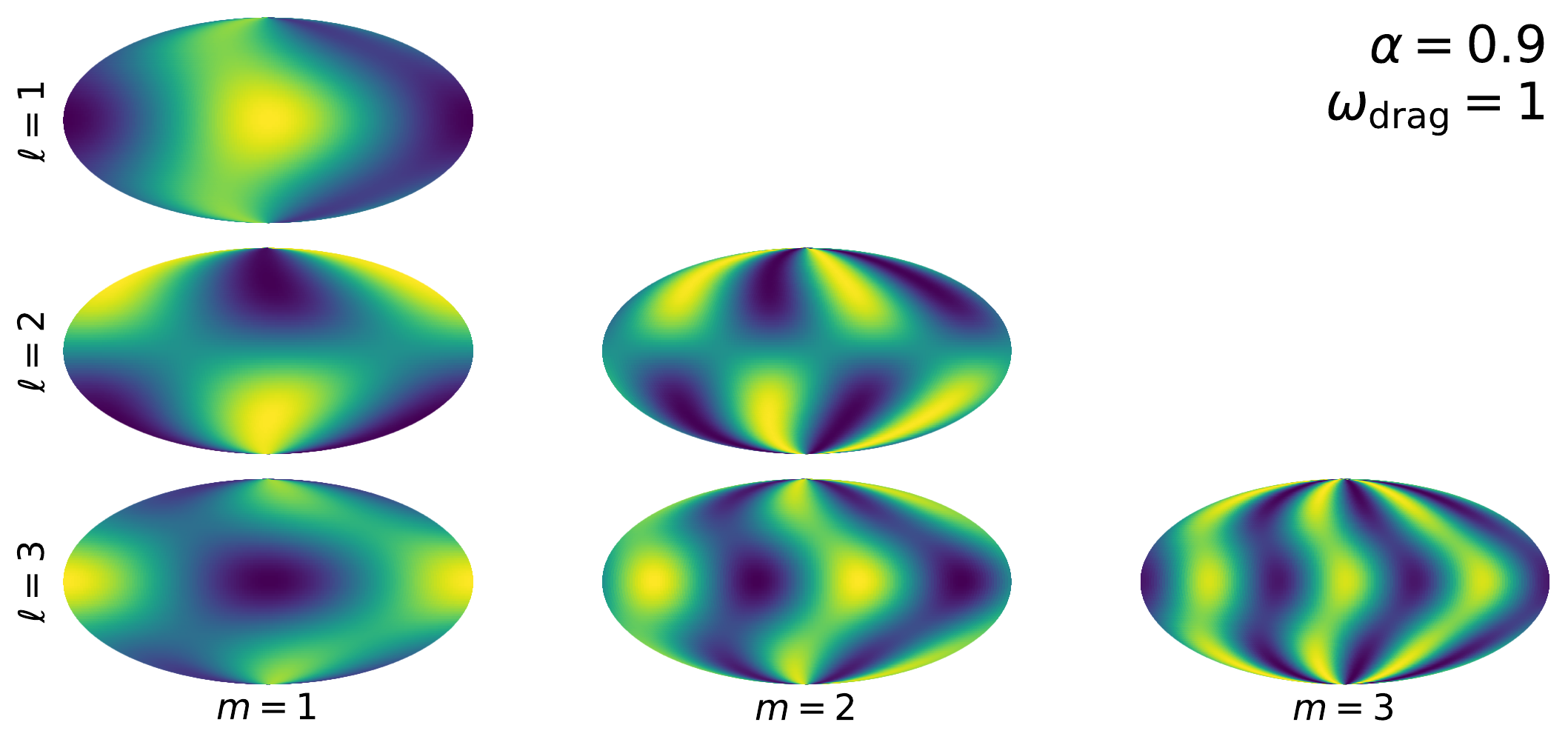}    \includegraphics[width=\columnwidth]{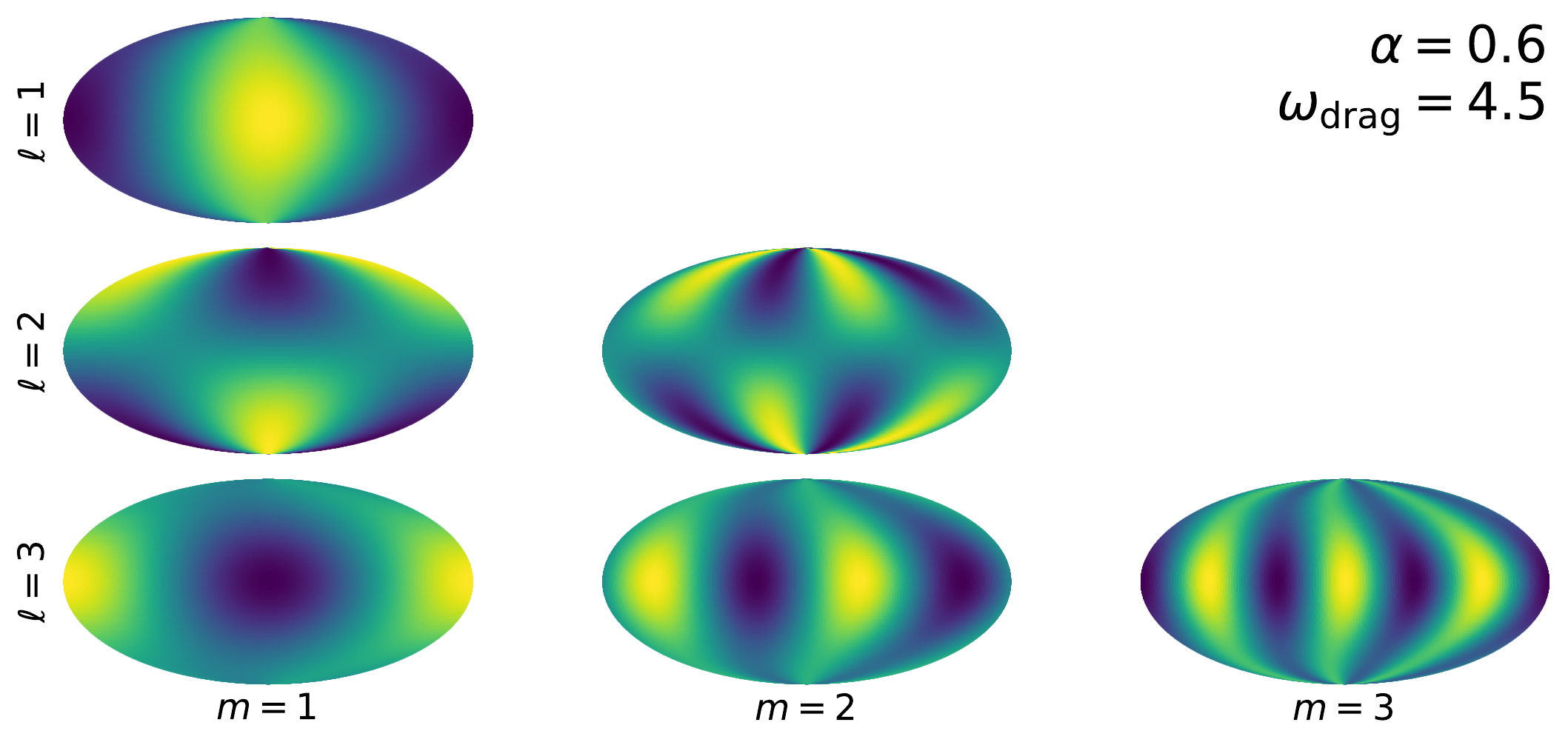}    \includegraphics[width=\columnwidth]{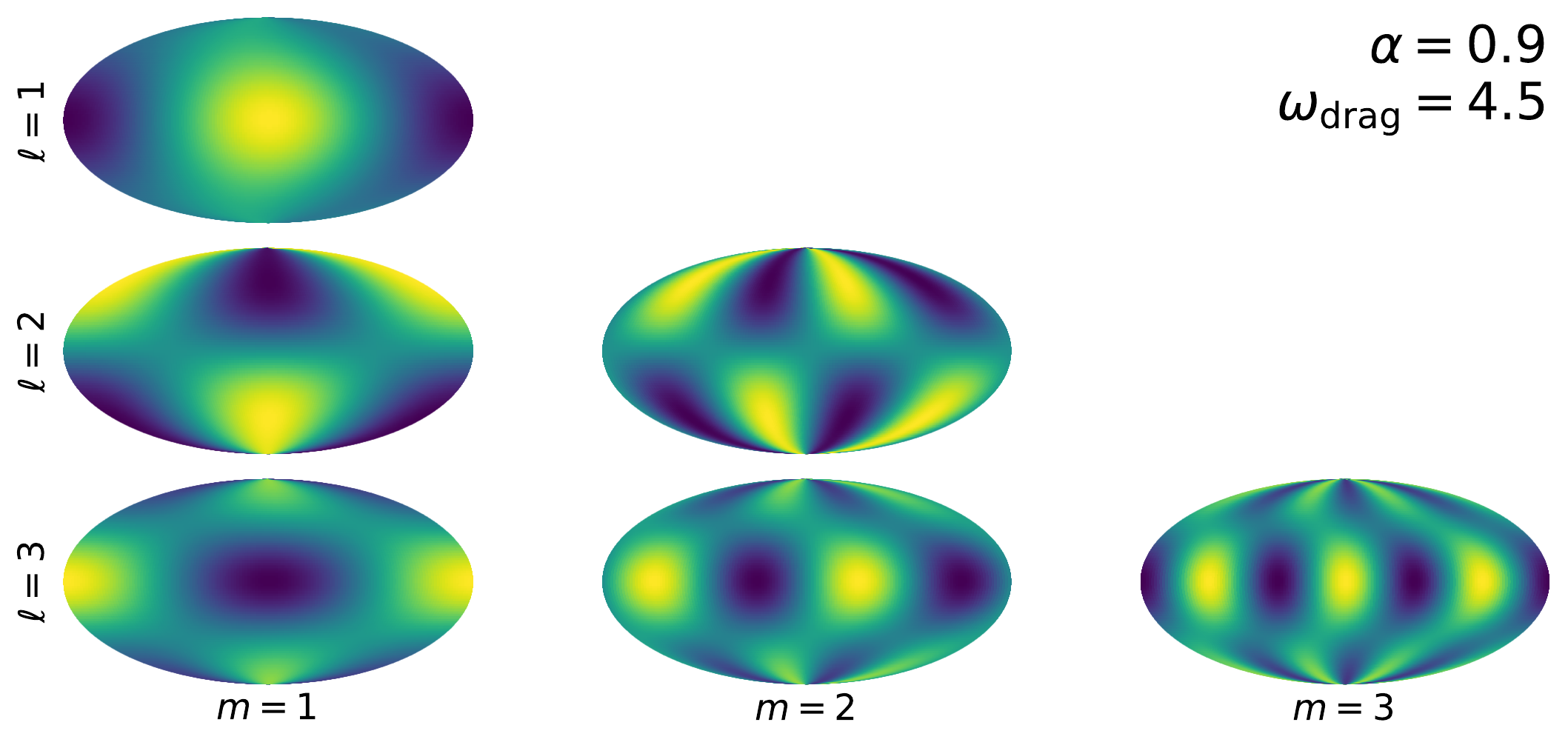}
    \caption{Four montages of the first several terms in the generalized spherical harmonic expansion of the temperature map in the \hml basis. Each montage corresponds to a different combination of $\alpha$ and $\omega_\mathrm{drag}$, labeled in the upper right of each quadrant. Each map shows the temperature perturbation (purple to yellow is cold to hot) as a function of latitude and longitude (shown in Mollweide projections such that the substellar longitude is in the center of the plot). The $m = 0$ terms are always zero. The even-$\ell$ terms are asymmetric about the equator and therefore do not represent typical GCM results in equilibrium, so we keep all even-$\ell$ power coefficients fixed to zero in the subsequent fits \citep[this approach assumes edge-on orbits, for exceptions see][]{Cowan2017}. In the upper two quadrants, when $\omega_\mathrm{drag}$ is set to a smaller value, the chevron shape becomes more pronounced. In the two quadrants on the right, when the $\alpha$ is set to a larger value, the concentration of heat near the equator is more pronounced. }
    \label{fig:ml}
\end{figure*}

The generalised spherical harmonic \hml basis defines an ``alphabet'' with only a few free parameters which we can use to succinctly describe exoplanet temperature fields, shown in Figure~\ref{fig:ml}. Each panel represents an individual generalized spherical harmonic expansion term, and we reconstruct temperature maps of exoplanets by evaluating linear combinations of these basis maps. 

We have fully defined the temperature map of the planet with the parameters above, and now we must construct the Planck function $\mathcal{B}_\lambda(T(\theta,\phi))$, integrate $\mathcal{B}_\lambda$ within the bandpass of observations, and then disk-integrate to construct a thermal phase curve. We integrate the ratio of flux from the planet and the host star \citep{Cowan2011}: 
\begin{equation}
    \frac{F_p}{F_\star} = \frac{1}{\pi I_\star} \left(\frac{R_p}{R_\star}\right)^2 \int_0^\pi \int_{-\xi-\pi/2}^{-\xi+\pi/2} I(\theta, \phi) \cos(\phi+\xi) \sin^2(\theta) ~d\phi ~d\theta \label{eqn:diskint}
\end{equation}
where the intensity $I$ is given by
\begin{equation}
I = \int \lambda \mathcal{F}_\lambda \mathcal{B}_\lambda(T(\theta, \phi)) d\lambda
\label{eq:intensity}
\end{equation}
for a filter response function $\mathcal{F}_\lambda$. The above relation neglects any reflected light. We assume the spectrum of each host star is accurately described by \textsf{PHOENIX} model spectra with the nearest effective temperature and surface gravity \citep{Husser2013}.

We have developed an efficient Python package for computing this triple integral for time series phase curve observations, written with \textsf{Cython}, \textsf{theano} and \textsf{JAX} for maximum speed and readability, called \textsf{kelp}, which is publicly available\footnote{\url{https://github.com/bmorris3/kelp}}.

\section{From 0D to 2D: Bond albedos and redistribution efficiencies} \label{sec:albedos}
\subsection{Background}
\label{sect:background}

Zero-dimensional (0D) ``box models" are used ubiquitously in exoplanet science.  As a prime example, the ``equilibrium temperature" of an exoplanet is standard fare in publications (e.g. \citealt{Seager2010}).  It originates from considering the stellar irradiation received by, and subsequently re-radiated over $\chi \pi$ steradians of, a spherical exoplanet.  The stellar luminosity is $4 \pi R_\star^2 \sigma_{\rm SB} T_\star^4$ with $R_\star$ being the stellar radius and $\sigma_{\rm SB}$ the Stefan-Boltzmann constant.  The star is approximated as a spherical blackbody with a uniform photospheric temperature $T_\star$.  The stellar flux fills a celestial sphere with a surface area of $4\pi a^2$, where $a$ is the distance between the star and exoplanet.  The exoplanet presents a cross section of $\pi R^2$ for intercepting this stellar flux, but absorbs only a fraction $1-A_{\rm B}$ of it with $A_{\rm B}$ being its Bond albedo.  Therefore, the box model of luminosity received is
\begin{equation}
L = 4 \pi R_\star^2 \sigma_{\rm SB} T^4_\star \left(1 - A_{\rm B} \right) \frac{\pi R^2}{4 \pi a^2}.
\label{eq:L}
\end{equation}
Assuming the exoplanet to radiate like a blackbody, the box model of luminosity re-radiated by the exoplanet is
\begin{equation}
L^\prime = \chi \pi R^2 \sigma_{\rm SB} T^4,
\label{eq:Lp}
\end{equation}
where $T$ is a characteristic temperature associated with the exoplanet.  If we assume that its flux is re-radiated over $\chi \pi = 4\pi$ steradians, then one recovers the standard expression for the equilibrium temperature (e.g., \citealt{Seager2010}),
\begin{equation}
T_{\rm eq} = \frac{T_{\rm irr}}{\sqrt{2}} \left(1 - A_{\rm B} \right)^{1/4},
\end{equation}
where $T_{\rm irr} \equiv T_\star \sqrt{R_\star/a}$ is the irradiation temperature.  The equilibrium temperature is often quoted assuming $A_{\rm B}=0$ (e.g., \citealt{Lopez-Morales2007}).

More generally, one may write $f = 1/\chi$ as a redistribution factor and derive an expression for the hemispherically-averaged dayside temperature of the exoplanet,
\begin{equation}
T_d = T_0 f^{1/4},
\label{eq:dayside0}
\end{equation}
where we define\footnote{Note that this definition includes the Bond albedo, which is different from \citet{Cowan2011b} who use $T_0 \equiv T_{\star}\sqrt{R_\star/a}$ to represent the irradiation temperature.} $T_0 \equiv T_{\rm irr} (1 - A_{\rm B})^{1/4}$.  By analogy, one may write the hemispherically-averaged nightside temperature as $T_n = T_0 g^{1/4}$.  By assuming a linear relationship between $f$ and $g$ and applying the boundary conditions that $f=g=1/4$ for full heat redistribution and $f=2/3$ and $g=0$ for no redistribution, one obtains
\begin{equation}
T_n = T_0 \left( \frac{2}{5} - \frac{3f}{5} \right)^{1/4}.
\label{eq:nightside0}
\end{equation}
One may also rewrite the preceding expressions in terms of a redistribution efficiency $\varepsilon = 8/5 - 12f/5$ that is bounded between 0 and 1, which yields equations 4 and 5 of \cite{Cowan2011b}.

The dayside and nightside brightness temperatures of an exoplanet may be directly inferred from data.  If one observes the secondary eclipse of an exoplanet purely in thermal emission, then the dayside brightness temperature may be extracted in a model-independent way \citep{Seager2010}.  Extracting the nightside brightness temperature requires a thermal phase curve to be measured as it is the difference in flux between the top of the transit and the bottom of the secondary eclipse \citep{Schwartz2017}.  For example, the nightside brightness temperature of 55 Cancri e was measured using a Spitzer phase curve \citep{Demory2016b}.  It is worth noting that, in extracting the temperature map, equation \ref{eq:intensity} already accounts for thermal emission within a \spitzer bandpass.  For a blackbody, the temperature and brightness temperature are equal.

Equations \ref{eq:dayside0} and \ref{eq:nightside0} may be inverted to solve for the Bond albedo and redistribution factor/efficiency, if the dayside and nightside (brightness) temperatures are known empirically.  However, it is possible to obtain negative Bond albedos by using equations \ref{eq:dayside0} and \ref{eq:nightside0}.  This artefact arises from approximating a two-dimensional (2D) situation with a 0D box model and is tied to the choice of boundary condition made for $g$, which we will next demonstrate.  It affects the inference of $A_{\rm B}$ and $f$ or $\varepsilon$ from all thermal phase curves of exoplanets.  

\subsection{General box models}

Starting again from $T_n = T_0 g^{1/4}$, we apply more general boundary conditions on $g$:
\begin{equation}
\begin{split}
\mbox{Full redistribution: } &~f = g = \frac{1}{4} , \\
\mbox{No redistribution: } &~f = f_0, ~g = 0.
\end{split}
\end{equation}
This yields a general expression for $g$ in terms of $f$ and $f_0$,
\begin{equation}
g = \frac{f-f_0}{1-4f_0}.
\end{equation}

Similarly, more general boundary conditions on $\varepsilon$ are:
\begin{equation}
\begin{split}
\mbox{Full redistribution: } &~f = \frac{1}{4}, ~\varepsilon=1, \\
\mbox{No redistribution: } &~f = f_0, ~\varepsilon=0,
\end{split}
\end{equation}
which yields
\begin{equation}
\varepsilon = \frac{4\left( f-f_0 \right)}{1-4f_0} = 4g.
\end{equation}

The general expression for the dayside temperature is
\begin{equation}
T_d = T_0 f^{1/4} = T_0 \left( \frac{\varepsilon}{4} - \varepsilon f_0 + f_0 \right)^{1/4}.
\label{eq:dayside}
\end{equation}
The general expression for the nightside temperature is
\begin{equation}
T_n = T_0 \left( \frac{f-f_0}{1-4f_0} \right)^{1/4} = \frac{T_0 \varepsilon^{1/4}}{\sqrt{2}}.
\label{eq:nightside}
\end{equation}
In the limit of $f_0=2/3$ \citep{Hansen2008}, we recover equation 4 of \cite{Cowan2011b} for the dayside temperature.

If we define the contrast between the nightside and dayside hemispheric temperatures as $C \equiv T_n/T_d$, then the redistribution factor may be expressed as
\begin{equation}
f = \frac{f_0}{1 - C^4 \left( 1 - 4f_0 \right)}.
\end{equation}

\subsection{Why box models may produce negative Bond albedos}

\begin{figure}[!h]
\begin{center}
\vspace{-0.1in}
\includegraphics[width=\columnwidth]{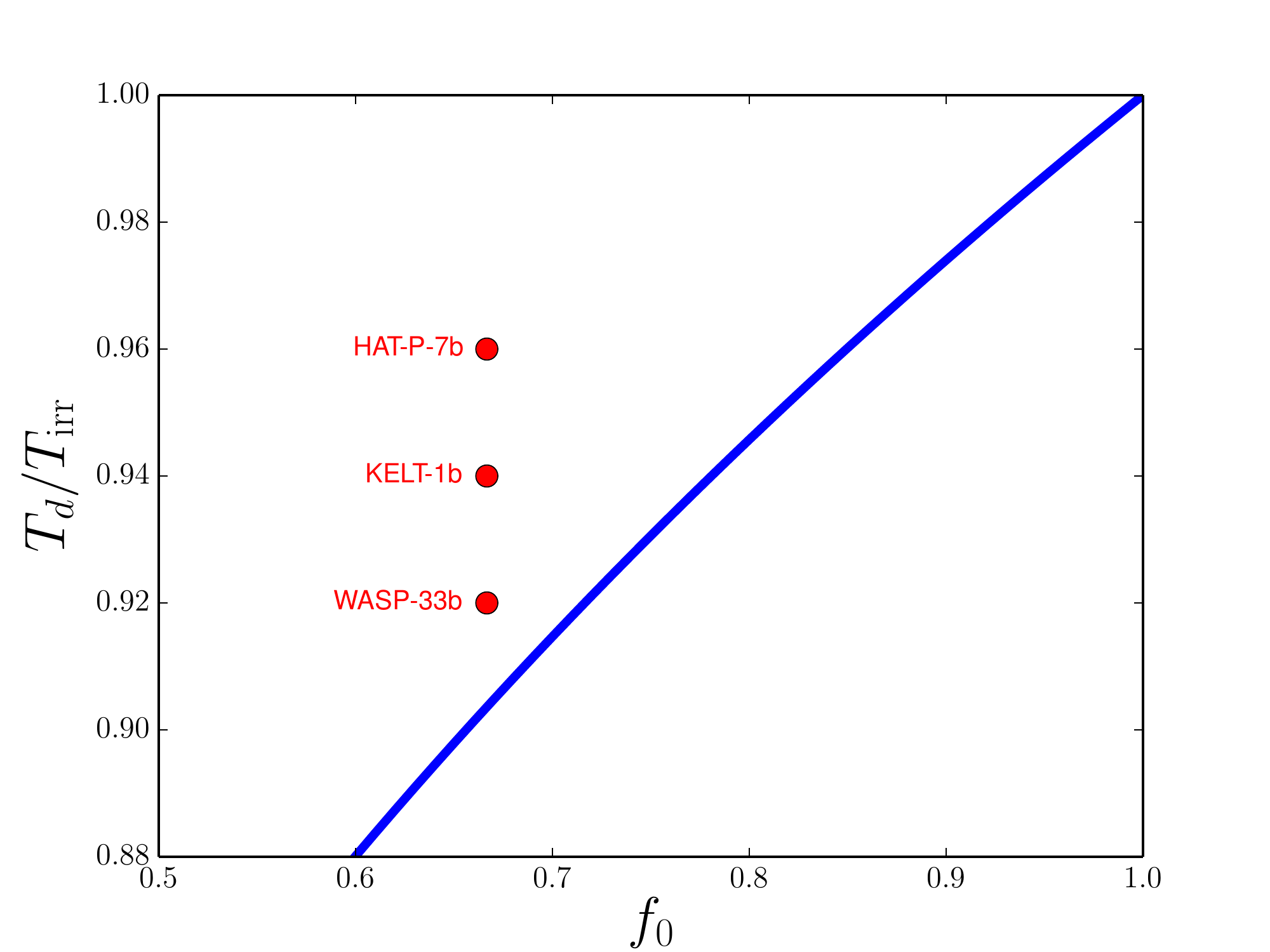}
\end{center}
\vspace{-0.2in}
\caption{Lower limit on $T_d/T_{\rm irr}$ (curve) for producing negative Bond albedos.  For comparison, empirically obtained values of $T_d/T_{\rm irr}$ from \cite{Wong2021} are shown; the uncertainties on these values (not shown for plot clarity) are between 2\% and 5\%.}
\label{fig:contrast}
\end{figure}

One may rewrite the expression for the dayside temperature in equation \ref{eq:dayside} to obtain an expression for the Bond albedo,
\begin{equation}
A_{\rm B} = 1 - \frac{1}{f} \left( \frac{T_d}{T_{\rm irr}} \right)^4.
\end{equation}
If we demand that $A_{\rm B}<0$, then we obtain the following inequality,
\begin{equation}
\frac{T_d}{T_{\rm irr}} > f^{1/4}.
\label{eq:lowerlimit}
\end{equation}
In the high-temperature limit, we expect $C \approx 0$ and $f \approx f_0$.  Therefore, negative Bond albedos are obtained when $T_d/T_{\rm irr} \gtrsim f_0^{1/4}$.

It only remains to derive the value of $f_0$.  The standard value of $f_0=2/3$ is obtained from adding an irradiation term to the expression for the intensity associated with Milne's solution \citep{Hansen2008}.  Section 2.2 of \citet{Cowan2011b} discusses how other choices may be made.  In the context of our box models, we may write
\begin{equation}
L^\prime = R^2 \int^{\pi/2}_{-\pi/2} \int^{\pi}_0 F \sin\theta ~d\theta ~d\phi,
\end{equation}
where the integration from $\phi = -\pi/2$ to $\pi/2$ occurs only over the dayside hemisphere.  If the dayside flux of the exoplanet is $F= \sigma_{\rm SB} T^4$ and $T$ is assumed to be constant, then we recover $L^\prime = 2 \pi R^2 \sigma_{\rm SB} T^4$, $\chi = 2$ and $f_0=1/2$.  If we account approximately for anisotropic emission over the dayside hemisphere (caused by anisotropic stellar irradiation; \citealt{Hansen2008}) by writing $F= \sigma_{\rm SB} T^4 \sin\theta$, then we obtain $f_0 = 2/\pi \approx 0.64$.  This assumes that the radiative timescale is much shorter than the dynamical timescale \citep{Lopez-Morales2007}.  

Table 8 of \cite{Wong2021} lists the dayside and irradiation temperatures of several hot Jupiters obtained from analysing Spitzer data.  Specifically, it allows us to estimate $T_d/T_{\rm irr}$ values of $0.96 \pm 0.04$ for HAT-P-7b, $0.94 \pm 0.05$ for KELT-1b, $0.92 \pm 0.02$ for WASP-18b and $0.83 \pm 0.02$ for WASP-33b.  Figure \ref{fig:contrast} demonstrates that these estimates of $T_d/T_{\rm irr}$ satisfy the condition for negative Bond albedos stated in equation \ref{eq:lowerlimit}.  Unsurprisingly, Table 8 of \cite{Wong2021} lists a negative Bond albedo for HAT-P-7b.

Generally, if the empirically-obtained value of $T_d/T_{\rm irr}$ exceeds about 0.9 (with this exact threshold depending on the assumed value of $f_0$), then the use of equations \ref{eq:dayside0} and \ref{eq:nightside0} results in a negative value for the Bond albedo.  This artifact originates from the global non-conservation of energy.

\subsection{General expression for Bond albedo} \label{sec:general-A_B}

We propose more general expressions for extracting $A_{\rm B}$ and $\varepsilon$ directly from a two-dimensional temperature map, which have been previously proposed in \citet{Keating2019}. Generally, the expression for $L$ in equation \ref{eq:L} holds regardless of one's assumption on $f_0$.  However, the expression for $L^\prime$ in equation \ref{eq:Lp} needs to be generalised for application to infrared phase curves,
\begin{equation}
L^\prime = R^2 \int^{\pi}_{-\pi} \int^{\pi}_0 F \sin\theta ~d\theta ~d\phi.
\end{equation}
A general expression for inference of the Bond albedo is
\begin{equation}
A_{\rm B} = 1 - \left( \frac{a}{R_\star} \right)^2 \frac{\int^{\pi}_{-\pi} \int^{\pi}_0 F \sin\theta ~d\theta ~d\phi}{\pi \sigma T_\star^4}.
\end{equation}
The two-dimensional flux distribution $F(\theta,\phi)$ may be inferred from an infrared phase curve, where the temperature $T(\theta,\phi)$ is not assumed to be constant.

The redistribution efficiency is simply the ratio of the nightside to dayside fluxes,
\begin{equation}
\varepsilon = \frac{\int^{\pi}_{\pi/2} \int^{\pi}_{0} F \sin\theta ~d\theta ~d\phi + \int^{-\pi/2}_{-\pi} \int^{\pi}_{0} F \sin\theta ~d\theta ~d\phi}{\int^{\pi/2}_{-\pi/2} \int^{\pi}_0 F \sin\theta ~d\theta ~d\phi}. \label{eqn:varepsilon2d}
\end{equation}

\subsection{Limitations of two-dimensional models}

While more general than 0D box models, the 2D temperature-map approach has the following limitations:
\begin{itemize}
    
    \item The atmospheric temperature generally varies in all three dimensions, implying that thermal phase curves measured at different wavelengths sample radial/vertical variations in temperature \citep{Stevenson2014}.  In other words, a thermal phase curve at a specific wavelength probes a 2D surface; at other wavelengths, other surfaces are probed. These surfaces are generally not isobaric.
    
    \item The spectral energy distribution of the exoplanet atmosphere may deviate strongly from a blackbody, especially far away from the blackbody peak where the intensity is proportional to $\lambda^{-4}$ (Rayleigh-Jeans law).  This implies that the brightness temperature is not the true local temperature.
    
\end{itemize}

In the limit of sparse data, it is plausible to combine the analysis of infrared phase curves, measured in two different Spitzer channels, to infer a 2D temperature map.  In the era of the James Webb Space Telescope (JWST), the need for a three-dimensional approach will be urgently needed as it will affect the inference of Bond albedo values from JWST phase curves.

\section{2D Temperature Maps from GCMs} \label{sec:gcms}

The incentive is strong to infer 2D temperature maps for exoplanets, despite the lack of constraints from phase curve observations. Planets such as KELT-9 b have dayside atmospheres as hot as a K star and nightside temperatures as hot as an M star, and the distribution of temperature with latitude may be similarly extreme, spanning $\sim1500$ K from pole to equator (see below). These tremendous temperature variations suggest that different chemistry may be important within the same planetary hemisphere. For example, this has important implications in translating what we learn from the phase curve into expectations for observations at the planet's terminator, which is probed during transmission spectroscopy.

The disk-integration that occurs when constructing a phase curve in Equation~\ref{eqn:diskint} effectively discards information about the distribution of temperatures with latitude. In some cases this information can be recovered for the daysides of exoplanets on slightly inclined orbits during secondary eclipse \citep[see, e.g.:][]{Majeau2012,deWit2012}. In general, however, the distribution of temperatures with latitude is not constrained by the phase curve. 

In this section, we explore whether we can use three-dimensional general circulation models to inform inferred temperature maps of real hot Jupiters from phase curve observations. We introduce a set of general circulation models (GCMs) in the next subsection, then outline a procedure for fitting them in the following subsection. Finally, we summarize the results in the limit of a planet with very high irradiation.

\subsection{GCM Simulations}

\begin{table}
    \centering
    \begin{tabular}{c|cc|c}
    Model & \multicolumn{2}{c}{$T_\mathrm{init}$ [K]} & \\
    Name & $\gamma = 0.5$ & $\gamma=2$ & $T_{\rm eq}$ [K] \\ \hline
    C &  589 & 501 & 545 \\
    C1 & 741 & 630 & 685 \\
    C2 & 806 & 686 & 745 \\
    W  & 1048 & 892 & 969 \\
    W1 & 1317 & 1121 & 1218 \\
    W2 & 1433 & 1219 & 1325 \\
    H & 1863 & 1585 & 1723 \\
    H1 & 2343 & 1994 & 2167 \\
    \end{tabular}
    \caption{Initial and equilibrium temperatures of the GCM runs of \citet{Perna2012}.}
    \label{tab:models}
\end{table}

Given that phase curves of exoplanets do not constrain the latitudinal distribution of heat in an exoplanet atmosphere, we will ``observe'' some ubiquitous trends in model atmospheres which set the expectations for the latitude distribution of heat in real exoplanet atmospheres. \citet{Perna2012} studied heat redistribution and energy dissipation in hot Jupiter atmospheres using three-dimensional circulation models with dual-band or ``double grey" radiative transfer. The authors used the GCMs to infer the effects of temperature inversions and stellar irradiation on hotspot offsets from the substellar point, among other quantities. We summarize their grid of 16 GCMs in Table~\ref{tab:models}.  The greenhouse parameter $\gamma$ is the ratio of the mean infrared to optical/visible opacity and controls whether a temperature inversion is present \citep{Guillot2010}; $\gamma=0.5$ and 2 correspond to temperature-pressure profiles without and with inversions, respectively.

\subsection{Fits to the GCM Temperature Maps}

\begin{figure*}
    \centering
    \includegraphics[width=\textwidth]{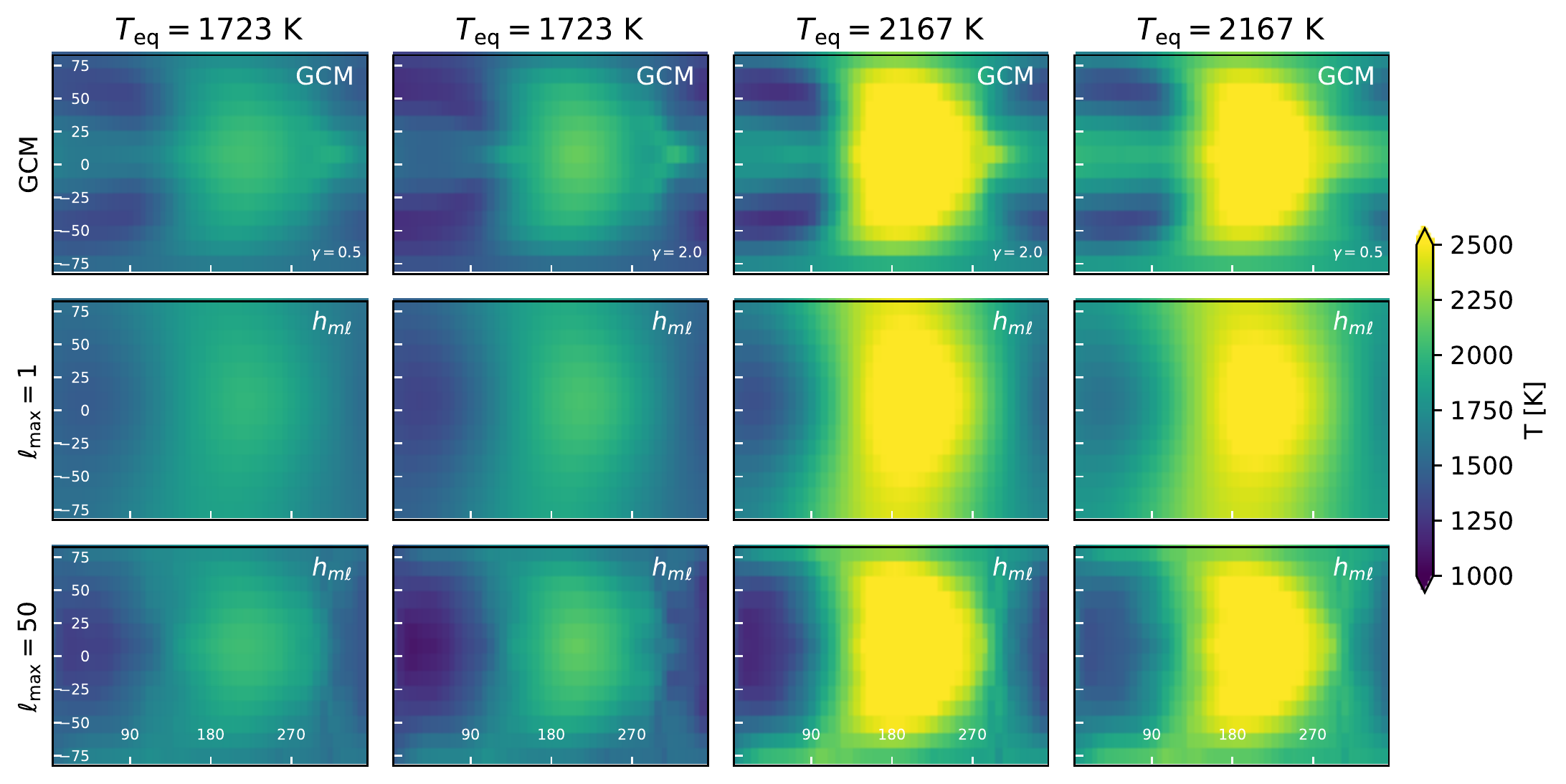}
    \caption{Comparison between the 2D temperature maps of hot Jupiter GCMs  in the upper row \citep{Perna2012},  and the 2D temperature maps represented with the \hml basis in the middle and lower rows with $\lmax=1$ and 50, respectively. The horizontal axis is longitude, the vertical axis is latitude, and the center of each map is the substellar point. For model input parameters see Table~\ref{tab:models}; for an outline of the linear algorithm used to fit maps with $\lmax=50$, see Appendix~\ref{app:a}.}
    \label{fig:hottest_maps}
\end{figure*}

\subsubsection{Phase offset}

The temperature map formulation by \citet{Heng2014a} is constructed without any special definitions for an origin in longitude, $\phi$. As a result, the temperature maps need a constant offset of $\pi/2$ to behave correctly in the limit of $\omega_\mathrm{drag} \gg 1$, which ensures that the substellar point is also the hottest point on the map. 

Similarly, we introduce an arbitrary phase offset as a fitting parameter, $\Delta\phi$, which allows the temperature map to be rotated with respect to the substellar point. While the \hml basis could in principle fit any temperature map at sufficiently large spherical harmonic degree, maps with signficant hotspot offsets would require $\lmax \gg 1$ in order to approximate the offset temperature distribution, incurring many free $C_{m\ell}$ parameters and making the inference intractable. Therefore, we introduce a single arbitrary rotation offset angle $\Delta\phi$ to the fits of the temperature maps to concisely describe hotspot offsets at small \lmax.

\subsubsection{Empirical limits on \lmax}

As noted in the previous section, introducing the coordinate system offset angle $\Delta\phi$ of the temperature map fits allows us to reproduce the GCM temperature maps with small \lmax. Choosing $\lmax=1$ requires five free parameters: $\{ f, C_{11}, \alpha, \omega_\mathrm{drag}, \Delta\phi \}$. The $\ell=2$ maps are not important for fitting mean phase curves of exoplanets as they are not symmetric across the planetary equator, and therefore should not have significant power for an atmosphere near equilibrium (see Figure~\ref{fig:ml}. $\lmax=3$ has seven parameters, and so on. We first fit the 2D GCM temperature maps with the \hml basis using $\lmax=3$, and find that the three $C_{ml}$ terms for $\ell=3$ are constrained to be very small compared to the power in $C_{11}$. We then inspect the quality of the fits for $\lmax=1$ and $\lmax=3$, and find that the $\lmax=1$ maps are indeed strikingly similar to the $\lmax=3$ maps, and thus we reason that $\lmax=1$ is likely sufficient to approximate temperature maps for phase curve analyses, requiring up to five free parameters. 

We note that previous studies have suggested that about five free parameters can be constrained by fits to a 1D phase curve \citep{Cowan2011,Hu2015}. It is encouraging that we arrive at a similar conclusion about the information content of phase curves as studies that take very different approaches to constructing phase curves, but as we will discuss in the next subsection, if we narrow our focus to the hottest exoplanets we can collapse the parameter space further.

\subsubsection{Trends towards high equilibrium temperature}

Next, we look for trends with irradiation in the physically meaningful fitting parameters with which we can constrain the 2D temperature maps of real exoplanets. We fit the 2D temperature maps of the \citet{Perna2012} GCMs for the parameters which control the 2D temperature map in the \hml basis with $\lmax=1$, $\{ f, C_{m\ell}, \alpha, \omega_\mathrm{drag}, \Delta\phi \}$, using Markov Chain Monte Carlo \citep{Foreman-Mackey2013}. The GCM temperature fields and the maximum-likelihood \hml maps are shown in Figure~\ref{fig:hottest_maps}. Comparison between the GCM results and the maximum-likelihood \hml reconstructions confirms that the major features of the map are reproduced by the \hml model at small \lmax, and more subtle features are reproduced at large \lmax. When $\alpha$ and $\omega_\mathrm{drag}$ are allowed to vary, the concentration of heat near the equator is easily reproduced by the \hml basis. 

\begin{figure}
    \centering
    \includegraphics[width=\columnwidth]{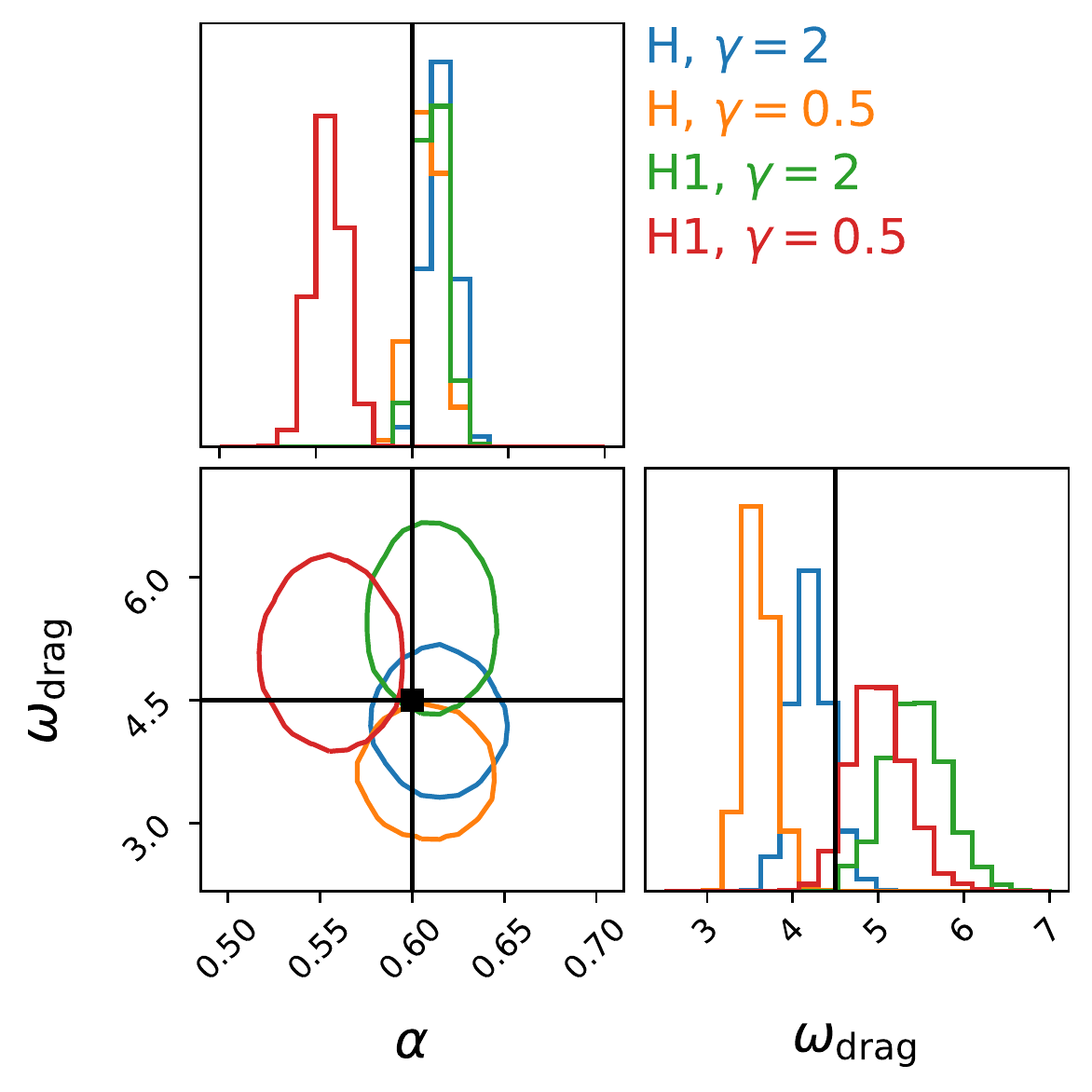}
    \caption{Posterior correlations between $\alpha$ and $\omega_\mathrm{drag}$ from fits of the \hml basis maps with $\lmax=1$ to the four hottest GCM temperature maps. The posterior distributions overlap at a few standard deviations for the values we adopt in the remainder of this work, $\alpha = 0.6$ and $\omega_\mathrm{drag} = 4.5$ (marked with black lines).}
    \label{fig:alpha_omega}
\end{figure}

Remarkably, the four hottest temperature maps in Figure~\ref{fig:hottest_maps} can all be fit with similar $\alpha$ and $\omega_\mathrm{drag}$ parameters, $\alpha=0.6$ and $\omega_\mathrm{drag} = 4.5$. This is demonstrated by the posterior distributions for $\alpha$ and $\omega_\mathrm{drag}$ for each of the four hottest GCM maps in Figure~\ref{fig:alpha_omega}. In other words, the latitudinal temperature distribution across maps with varying equilibrium temperature is quite similar when phrased in terms of the \hml ``alphabet''. The chevron shapes that are produced by this combination of $\alpha$ and $\omega_\mathrm{drag}$ will be familiar to many GCM practitioners -- it is a shape reproduced by many GCMs and has an elucidated theoretical basis \citep[e.g.:][]{Showman2011, Tsai2014}.

At $\lmax=3$ we are unable to reproduce some of the subtle features of the map, such as the equatorial jet that is noticeable in the nightside atmosphere, or the ``point'' that forms at the leading edge of the dayside atmosphere. These features could be reproduced by going to higher $\lmax$, but since we are focused here on jointly inferring properties from GCMs and phase curves, we assume that such fine features will remain undetectable with phase curve photometry, and neglect these small features. In the bottom row of Figure~\ref{fig:hottest_maps}, we show a linear \hml solution to the GCM temperature maps with $\lmax=50$, which better reproduces the precise shape of the dayside hotspot but not the jet, for example.

To summarize, our experiments fitting the \hml maps to the GCMs have shown that: (1) we must introduce a phase offset that was not accounted for in the original definition of the \hml basis; (2) $\lmax=1$ or 3 is likely sufficient to capture the global-scale temperature inhomogeneities that are routinely in agreement between different GCMs and likely to be constrained by the phase curves; (3) the hottest maps all demand a common combination of $\alpha=0.6$ and $\omega_\mathrm{drag} = 4.5$, collapsing our parameter space from five to three free parameters.

\section{\spitzer Observations} \label{sec:spitzer}

\begin{figure*}[ht!]
    \centering
    \includegraphics[width=\textwidth]{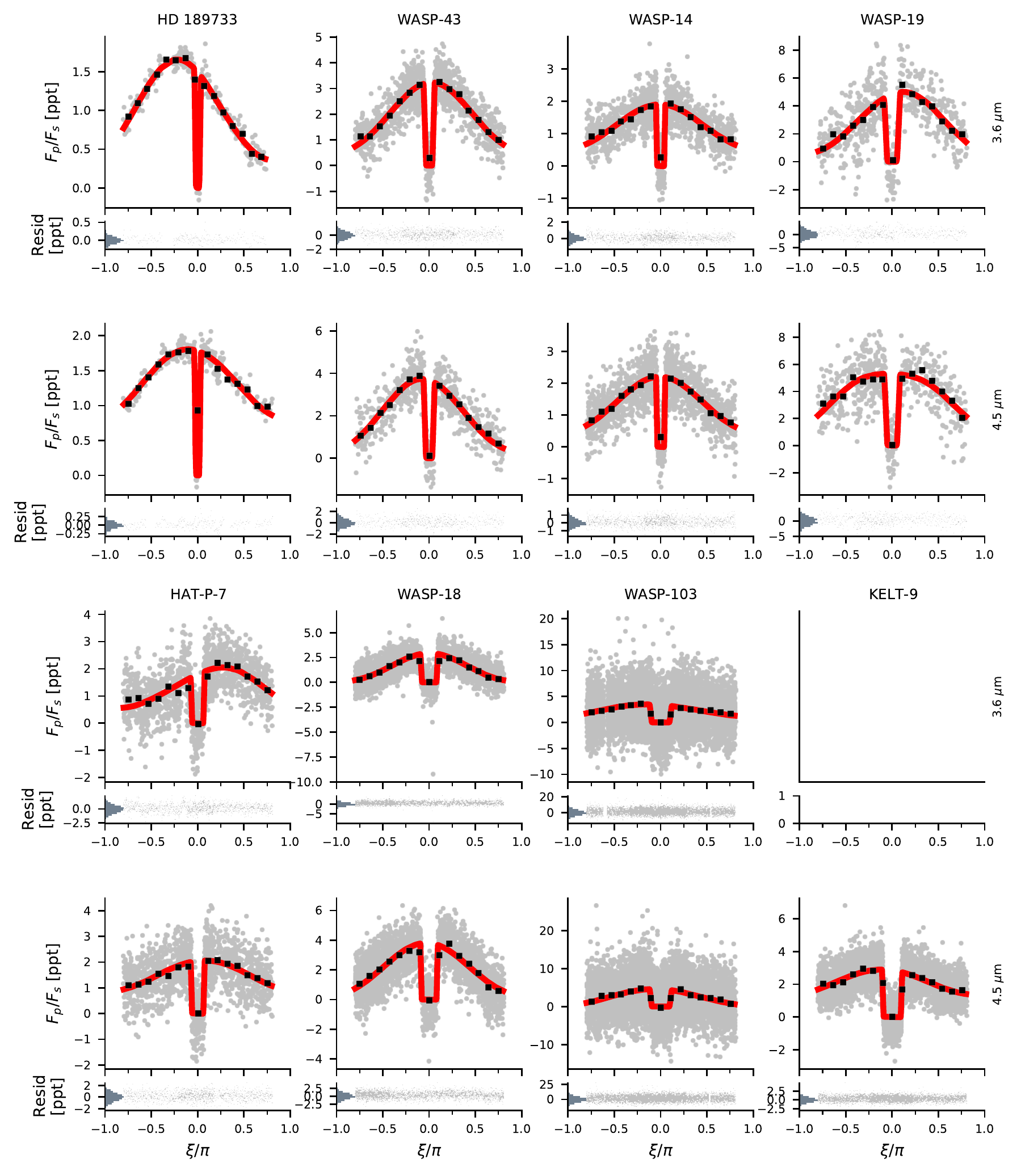}
    \caption{\spitzer phase curves of the eight exoplanets (gray, binned in black) fit with the \hml basis (red), plotted in order of equilibrium temperature (low to high from top left to bottom right). Each planet is plotted with the IRAC Channel 1 phase curve in the first and third row and the Channel 2 phase curve in the second and fourth row (see row labels on the right side). Maximum likelihood parameters for each fit are listed in Table~\ref{tab:spitzer}. Residuals are plotted beneath each fit with a histogram of residuals on the left.}
    \label{fig:spitzer_sequence}
\end{figure*}

\begin{table}
    \centering
    \begin{tabular}{lr}
        Planet & Reference \\ \hline
        HD 189733 & \citet{Knutson2012} \\
        WASP-18 & \citet{Maxted2013} \\
        WASP-14 & \citet{Wong2015}\\
        WASP-19 & \citet{Wong2016}\\
        HAT-P-7 & \citet{Wong2016}\\
        WASP-103 & \citet{Kreidberg2018}\\
        WASP-43 & \citet{Mendonca2018} \\
        KELT-9 & \citet{Mansfield2020} 
         \end{tabular}
    \caption{References for \spitzer phase curves presented in this work.}
    \label{tab:refs}
\end{table}

Next, we apply the \hml basis functions in the context of phase curve observations from the \spitzer Space Telescope. \spitzer photometry at 3.6 and 4.5 $\mu$m provides coverage of spectral regions where the majority of the planetary flux is thermal emission rather than reflected light (more on this in Section~\ref{sec:reflected}). The \spitzer phase curves are disk-integrated measurements of the planet-to-star flux ratio, and thus constrain the longitudinal distribution of flux from the planet. In this section, we present original photometric extractions for seven of the eight targets presented, and interpret those phase curves with the \hml basis.

\subsection{Reduction}

We retrieved archival \spitzer/IRAC observations from the \spitzer Heritage Archive\footnote{\url{http://sha.ipac.caltech.edu}}, as published in the references listed in Table~\ref{tab:refs}. The reduction of these AORs follows the procedure of \citet{Demory2016b}, and we refer the reader to details in that reference. We model the IRAC intra-pixel sensitivity \citep{Ingalls2016} using a modified implementation of the BLISS (BiLinearly-Interpolated Sub-pixel Sensitivity) mapping technique (\citealt{Stevenson2012}).

The baseline photometric model includes the pixel response function's FWHM along the $x$ and $y$ axes, which significantly reduces the level of correlated noise as shown in previous studies (e.g. \citealt{Lanotte2014}; \citealt{Demory2016}; \citealt{Demory2016b}; \citealt{Gillon2017}, \citealt{Mendonca2018}). The model does not include time-dependent parameters. We vary the sampling parameters with Markov Chain Monte Carlo (MCMC) in a framework already presented in \citet{Gillon2012}. We run two chains of 200,000 steps each to determine the phase-curve properties at 3.6 and 4.5 $\mu$m based on the entire dataset. We trimmed the ramp in the first 30 minutes of the first AOR of the HAT-P-7 Channel 1 visit.

The photometry of KELT-9 b in Channel 1 is being studied in a forthcoming work by Beatty et al.\ (in prep.), and therefore is excluded from this work. The photometry of HD 189733 is affected by stellar variability at the mmag level requiring a special analysis unique to this object, including time-dependent parameters which we exclude from the other analyses. As such, we adopt the fully-reduced \spitzer phase curve from \citet{Knutson2012} for subsequent analysis in both \spitzer channels.

\subsection{Analysis}

In the previous section we determined that the \hml basis can be reduced to three fitting parameters: the redistribution factor (or greenhouse parameter) $f$ or redistribution efficiency $\varepsilon$, the power in the first spherical harmonic term $C_{11}$, and the phase offset parameter $\Delta\phi$. In practice, these parameters respectively set: the baseline temperature field, the amplitude of the temperature variations from the day to the night side, and the hotspot offset. 

Each phase curve can be fit using the \hml basis with the free parameters: $\{C_{11}, \Delta\phi, \varepsilon\}$. We compute posterior distributions on each parameter with the No U-Turn Sampler using a \textsf{STAN}-like tuning schedule, implemented by \citep{Foreman-Mackey2021}. $\alpha$ and $\omega_\mathrm{drag}$ are varied with tight Gaussian priors: $\alpha \sim \mathcal{N}(0.6, 0.1)$ and $\omega_\mathrm{drag} \sim \mathcal{N}(4.5, 1.5)$. By allowing these parameters to have non-zero variance, we are slightly exploring the degeneracies between parameters, especially between $\alpha$ and $C_{11}$ which are highly correlated in fits to the phase curve with $\lmax=1$. We compute a fixed eclipse model for the planets using the best-fit planetary parameters from each paper in Table~\ref{tab:refs} using the \citet{Mandel2002} model and implemented by \citet{Kreidberg2015}. The eclipse model incurs no new free parameters -- we assume that all of the flux stemming from the phase curve signal vanishes during totality of secondary eclipse, so that the flux ratio $F_p/F_\star = 0$ in eclipse.

We allow the thermal phase curves to be contaminated by sinusoidal contributions from ellipsoidal variations and Doppler beaming with unknown amplitudes but fixed phases and periods. Typical amplitudes for these sinusoidal terms are small compared to the phase variations. The most significant ellipsoidal variations affected the phase curves of WASP-103 and WASP-19.

In Figure~\ref{fig:spitzer_sequence}, a typical draw from the posterior distribution is plotted in red for each of the \spitzer phase curves shown in gray, and binned in black. The upper row contains observations from the $3.6\,\rm\mu$m Channel 1, and the lower row contains $4.5\,\rm\mu$m Channel 2 observations. The phase curves are plotted in order of increasing equilibrium temperature from left to right. The fits in red are quite good approximations to the behavior of the light curves in black. A notable exception not shown here is WASP-12, which has been claimed to have a special phase curve shape at $4.5\,\rm\mu$m due to co-orbiting material between the planet and the star \citep{Bell2019}.

One notable phenomenon in Figure~\ref{fig:spitzer_sequence} is the variation in phase curve amplitude between \spitzer channels, particularly for WASP-19 b. The difference in amplitude and phase offset between the two channels is likely because different wavelengths probe different depths and therefore different temperatures in the planet's atmosphere. We discuss this more in Section~\ref{sec:discussion}.

\begin{figure}
    \centering
    \includegraphics[width=\columnwidth]{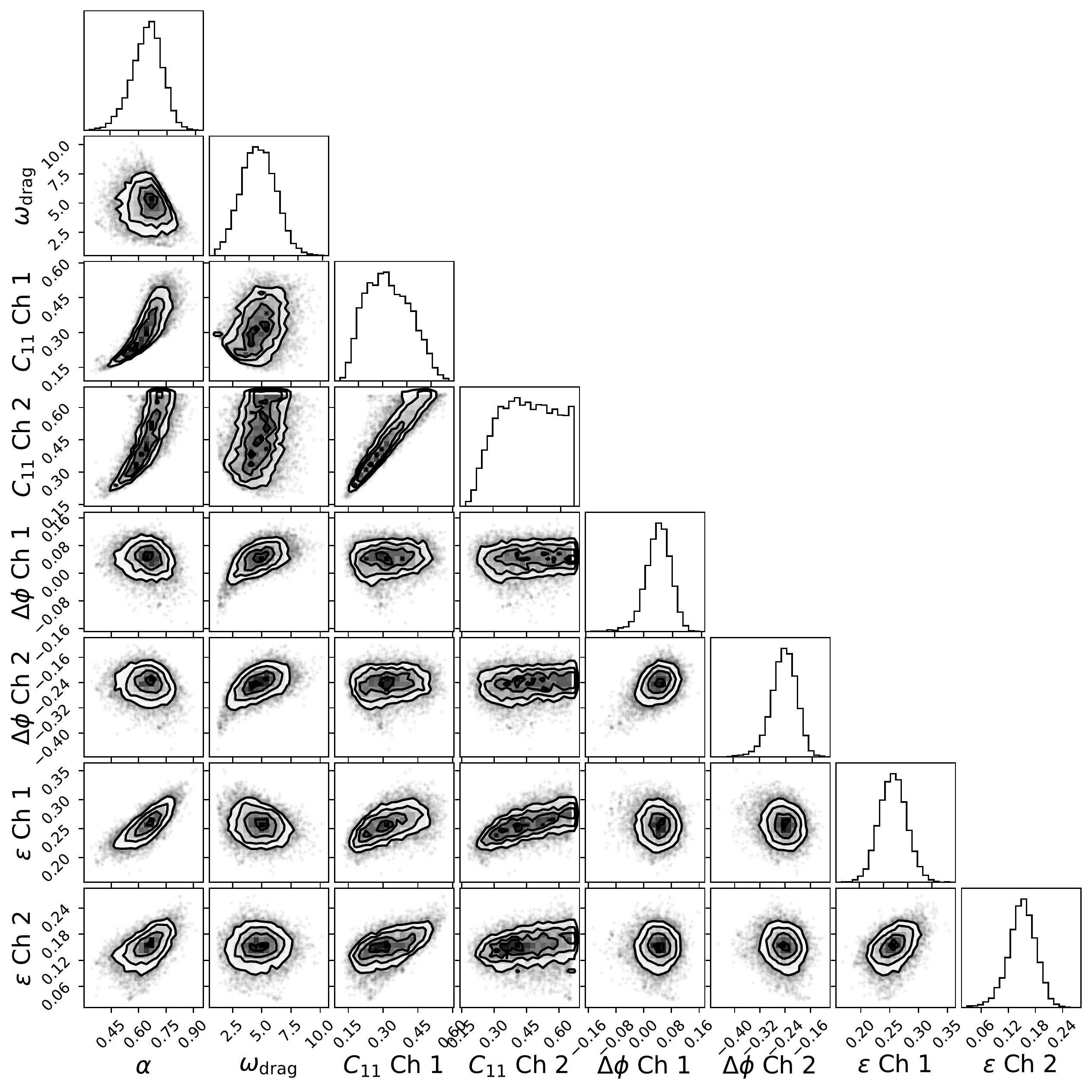}
    \caption{Posterior distributions for the three derived parameters $\{\Delta \phi, C_{11}, \varepsilon\}$ which determine the amplitude and shape of the phase curve for WASP-43 b as observed in \spitzer IRAC 1 and 2 (3.6, 4.5 $\rm \mu$m), with priors on $\{\alpha, \omega_\mathrm{drag}\}$. The global $\alpha$ parameter is degenerate with the $C_{11}$ specific to each bandpass, which introduces a cross-bandpass correlation between the two $C_{11}$ parameters.}
    \label{fig:wasp43}
\end{figure}

Posterior distributions and parameter correlations are shown for WASP-43 b in Figure~\ref{fig:wasp43}. The spherical harmonic power term $C_{11}$ is correlated with the dimensionless fluid parameter $\alpha$, which has a tight prior.

\begin{figure}
    \centering
    \includegraphics[width=\columnwidth]{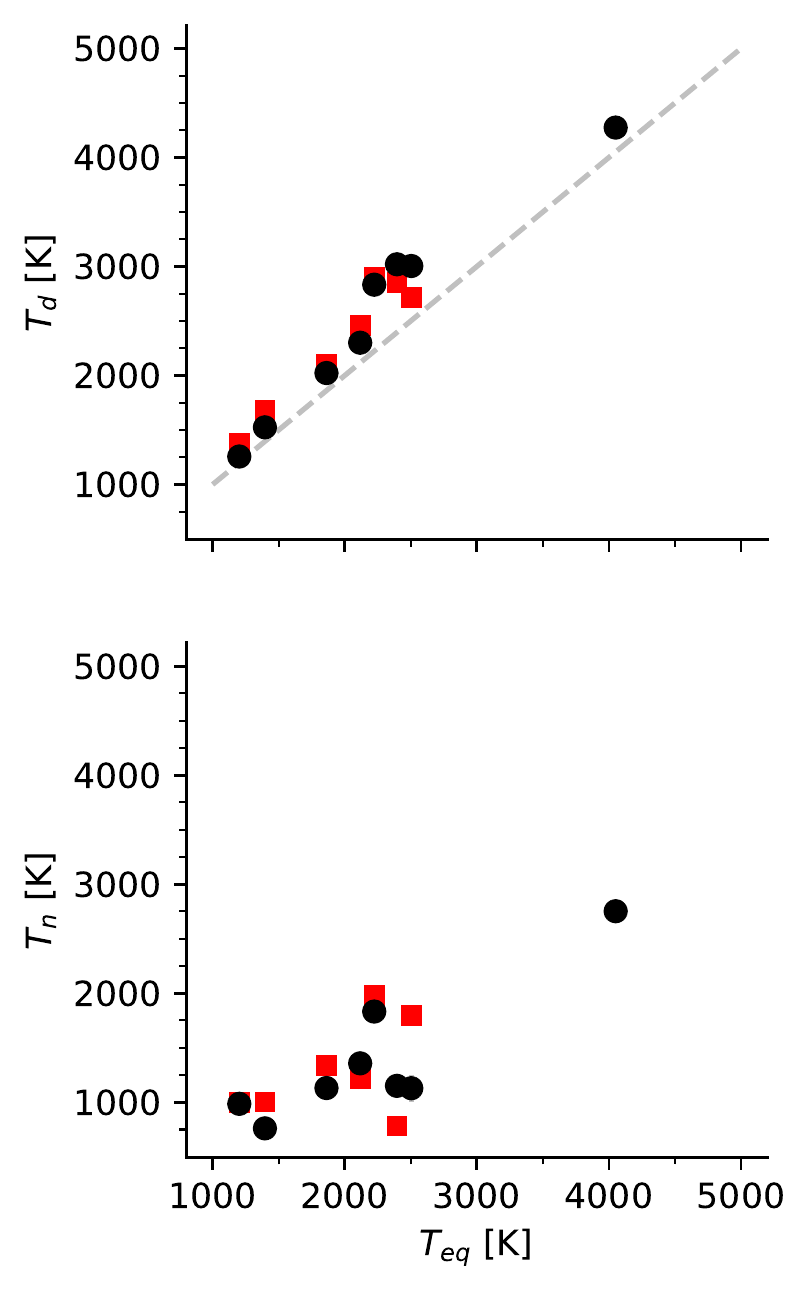}
    \caption{Hemispherically-averaged dayside and nightside temperatures $T_d, T_n$ in \spitzer Channel 1 (red squares) and Channel 2 (black circles) as a function of planetary equilibrium temperature. These temperatures are derived by averaging the dayside and nightside temperature maps derived from the full phase curve fit. The hottest planet is KELT-9 b, which clearly breaks the apparent ``uniform nightside temperature'' trend among cooler hot Jupiters noted by \citet{Keating2019}. Uncertainties are smaller than the points.
    \label{fig:temps_day_night}}
\end{figure}

Figure~\ref{fig:temps_day_night} shows the dayside and nightside temperatures derived from the \spitzer photometry of the hot Jupiters. We confirm that dayside temperatures are proportional to the equilibrium temperatures as found by \citet{Schwartz2015}. We also independently confirm the observation of \citet{Keating2019, Beatty2019} which suggested that the nightside temperatures of most hot Jupiters are similar to 1100 K, irrespective of equilibrium temperature. It has been suggested that silicates and hydrocarbon hazes can produce this trend \citep{Gao2020}. We note that this trend is broken by KELT-9 b, which is hot enough to dissociate H$_2$ molecules on its dayside, affecting the global heat transport \citep[see, e.g.:][]{Bell2018,Wong2020,Roth2021,Mansfield2020}.

\begin{figure}
    \centering
    \includegraphics[width=\columnwidth]{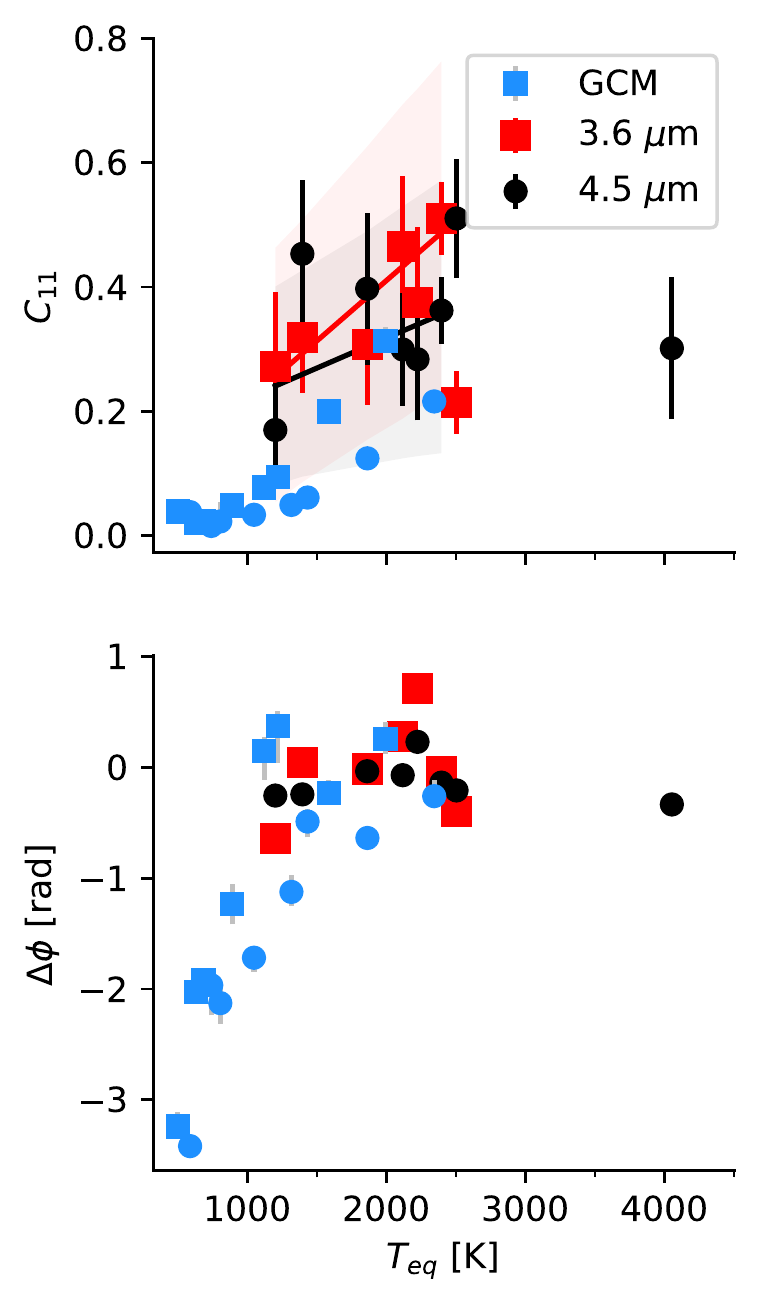}
    \caption{Maximum-likelihood parameters for the spherical harmonic power coefficient $C_{11}$ (upper) and hotspot offset (lower) for the full sample of GCM fits and \spitzer phase curve fits in two filter bands. Blue points are the results of fits to GCMs, circles and squares correspond to inverted and non-inverted atmospheres. Red squares and black circles represent fits to the \spitzer phase curves in filters IRAC 1 and 2, respectively.
    \label{fig:best_params}}
\end{figure}

The best-fit \hml basis parameters $C_{11}$ and $\Delta\phi$ for both the fits to the GCM temperature maps and the \spitzer phase curves are plotted in Figure~\ref{fig:best_params}. The $C_{11}$ values from the fits to the hottest GCMs agree well with the results from the \spitzer phase curves. The \spitzer phase curves show a roughly linear increase with equilibrium temperature given by
\begin{eqnarray}
C_{11, \mathrm{Ch 1}} &=& (0.00019 \pm 0.00008) T_{\rm eq} + (0.02 \pm 0.17),\\
C_{11, \mathrm{Ch 2}} &=& (0.00010 \pm 0.00007) T_{\rm eq} + (0.12 \pm 0.14),
\end{eqnarray}
for planetary equilibrium temperatures between 1200 K and 3100 K. The phase offset parameter $\Delta\phi$ is close to zero for all \spitzer phase curves except for the two most extreme temperature cases, while all GCMs below 1500 K indicate significantly nonzero phase offsets. The rather extreme phase offset of HD 189733 b is consistent with the range observed in GCM temperature maps of similar equilibrium temperatures. The simulations and the observations are in good agreement in terms of both $C_{11}$ and $\Delta\phi$ where the equilibrium temperatures overlap near 2000 K.

\renewcommand{\arraystretch}{1.5}

\begin{table*}
\centering
\begin{tabular}{cccccccccc}
Star & $\alpha$ & $\omega_\mathrm{drag}$ & \spitzer & $T_d$ & $T_n$ & $A_B$ & $\varepsilon$ & $\Delta \phi$ & $C_{11}$ \\
 &  & & Channel & [K]& [K] & & & [deg] & \\ \hline
HD 189733 & ${0.68}_{-0.10}^{+0.10}$ & ${5.02}_{-1.46}^{+1.45}$ & 1 & ${1373}_{-4}^{+3}$ & ${997}_{-15}^{+12}$ & ${-0.02}_{-0.02}^{+0.02}$ & ${0.41}_{-0.03}^{+0.03}$ & ${-36}_{-3}^{+3}$ & ${0.27}_{-0.09}^{+0.16}$ \\
 &  &  & 2 & ${1256}_{-3}^{+3}$ & ${985}_{-11}^{+9}$ & ${0.22}_{-0.01}^{+0.01}$ & ${0.52}_{-0.03}^{+0.03}$ & ${-14}_{-3}^{+5}$ & ${0.17}_{-0.05}^{+0.10}$ \\
WASP-43 & ${0.65}_{-0.08}^{+0.07}$ & ${4.63}_{-1.42}^{+1.39}$ & 1 & ${1679}_{-6}^{+6}$ & ${1001}_{-20}^{+18}$ & ${-0.02}_{-0.01}^{+0.02}$ & ${0.26}_{-0.02}^{+0.02}$ & ${2}_{-2}^{+2}$ & ${0.31}_{-0.08}^{+0.10}$ \\
 &  &  & 2 & ${1524}_{-10}^{+10}$ & ${760}_{-52}^{+37}$ & ${0.38}_{-0.02}^{+0.02}$ & ${0.15}_{-0.03}^{+0.03}$ & ${-14}_{-2}^{+2}$ & ${0.44}_{-0.12}^{+0.13}$ \\
WASP-14 & ${0.67}_{-0.09}^{+0.07}$ & ${4.88}_{-1.40}^{+1.39}$ & 1 & ${2098}_{-11}^{+11}$ & ${1336}_{-29}^{+28}$ & ${0.17}_{-0.02}^{+0.02}$ & ${0.31}_{-0.03}^{+0.03}$ & ${0}_{-4}^{+2}$ & ${0.31}_{-0.10}^{+0.11}$ \\
 &  &  & 2 & ${2022}_{-10}^{+10}$ & ${1130}_{-23}^{+21}$ & ${0.35}_{-0.01}^{+0.01}$ & ${0.22}_{-0.02}^{+0.02}$ & ${-2}_{-2}^{+2}$ & ${0.40}_{-0.13}^{+0.15}$ \\
WASP-19 & ${0.64}_{-0.09}^{+0.07}$ & ${4.61}_{-1.33}^{+1.45}$ & 1 & ${2457}_{-34}^{+35}$ & ${1214}_{-106}^{+75}$ & ${0.21}_{-0.04}^{+0.04}$ & ${0.16}_{-0.04}^{+0.04}$ & ${15}_{-2}^{+3}$ & ${0.46}_{-0.13}^{+0.13}$ \\
 &  &  & 2 & ${2300}_{-40}^{+38}$ & ${1356}_{-100}^{+97}$ & ${0.33}_{-0.06}^{+0.06}$ & ${0.24}_{-0.04}^{+0.05}$ & ${-4}_{-3}^{+4}$ & ${0.30}_{-0.08}^{+0.10}$ \\
HAT-P-7 & ${0.65}_{-0.10}^{+0.08}$ & ${4.77}_{-1.59}^{+1.33}$ & 1 & ${2903}_{-45}^{+49}$ & ${1975}_{-94}^{+80}$ & ${-0.67}_{-0.11}^{+0.11}$ & ${0.39}_{-0.05}^{+0.06}$ & ${40}_{-4}^{+5}$ & ${0.37}_{-0.12}^{+0.15}$ \\
 &  &  & 2 & ${2829}_{-33}^{+33}$ & ${1824}_{-53}^{+45}$ & ${-0.38}_{-0.05}^{+0.05}$ & ${0.30}_{-0.03}^{+0.03}$ & ${13}_{-5}^{+4}$ & ${0.20}_{-0.04}^{+0.04}$ \\
WASP-18 & ${0.50}_{-0.04}^{+0.06}$ & ${4.41}_{-1.41}^{+1.35}$ & 1 & ${2854}_{-20}^{+28}$ & ${781}_{-27}^{+35}$ & ${0.27}_{-0.03}^{+0.02}$ & ${0.03}_{-0.00}^{+0.01}$ & ${-2}_{-1}^{+2}$ & ${0.51}_{-0.05}^{+0.07}$ \\
 &  &  & 2 & ${3019}_{-28}^{+23}$ & ${1150}_{-68}^{+38}$ & ${0.01}_{-0.03}^{+0.04}$ & ${0.06}_{-0.01}^{+0.01}$ & ${-7}_{-2}^{+2}$ & ${0.36}_{-0.04}^{+0.06}$ \\
WASP-103 & ${0.59}_{-0.07}^{+0.07}$ & ${4.32}_{-1.55}^{+1.48}$ & 1 & ${2714}_{-39}^{+39}$ & ${1795}_{-85}^{+70}$ & ${0.25}_{-0.04}^{+0.05}$ & ${0.30}_{-0.04}^{+0.04}$ & ${-22}_{-4}^{+3}$ & ${0.21}_{-0.05}^{+0.05}$ \\
 &  &  & 2 & ${3004}_{-55}^{+55}$ & ${1128}_{-144}^{+121}$ & ${0.19}_{-0.05}^{+0.06}$ & ${0.07}_{-0.02}^{+0.03}$ & ${-11}_{-4}^{+2}$ & ${0.51}_{-0.11}^{+0.10}$ \\
KELT-9 & ${0.66}_{-0.10}^{+0.09}$ & ${4.76}_{-1.85}^{+1.50}$ & 1 & -- & -- & -- & -- & -- & -- \\
 &  &  & 2 & ${4273}_{-24}^{+23}$ & ${2751}_{-33}^{+33}$ & ${0.35}_{-0.01}^{+0.01}$ & ${0.30}_{-0.03}^{+0.03}$ & ${-19}_{-3}^{+3}$ & ${0.30}_{-0.09}^{+0.15}$ \\
\end{tabular}
\caption{Maximum-likelihood parameters for the fits to the \spitzer phase curves shown in Figure~\ref{fig:spitzer_sequence} with the priors $\alpha \sim \mathcal{N}(0.6, 0.1)$ and $\omega_\mathrm{drag} \sim \mathcal{N}(4.5, 1.5)$. The parameters are: the dimensionless fluid number $\alpha$, the dimensionless drag frequency $\omega_\mathrm{drag}$, the Spitzer IRAC channel corresponding to the remaining quantities (Channel 1 is centered on $3.6~\mu$m, Channel 2 is centered on $4.5~\mu$m), the integrated dayside temperature $T_d$, integrated nightside temperature $T_n$, the Bond albedo $A_B$, the heat redistribution efficiency $\varepsilon$, the phase offset $\Delta\phi$, and the power in the first spherical harmonic term $C_{11}$. Reported uncertainties assume the stellar effective temperature is known precisely.}
\label{tab:spitzer}
\end{table*}

\begin{figure*}
    \centering
    \includegraphics[width=\textwidth]{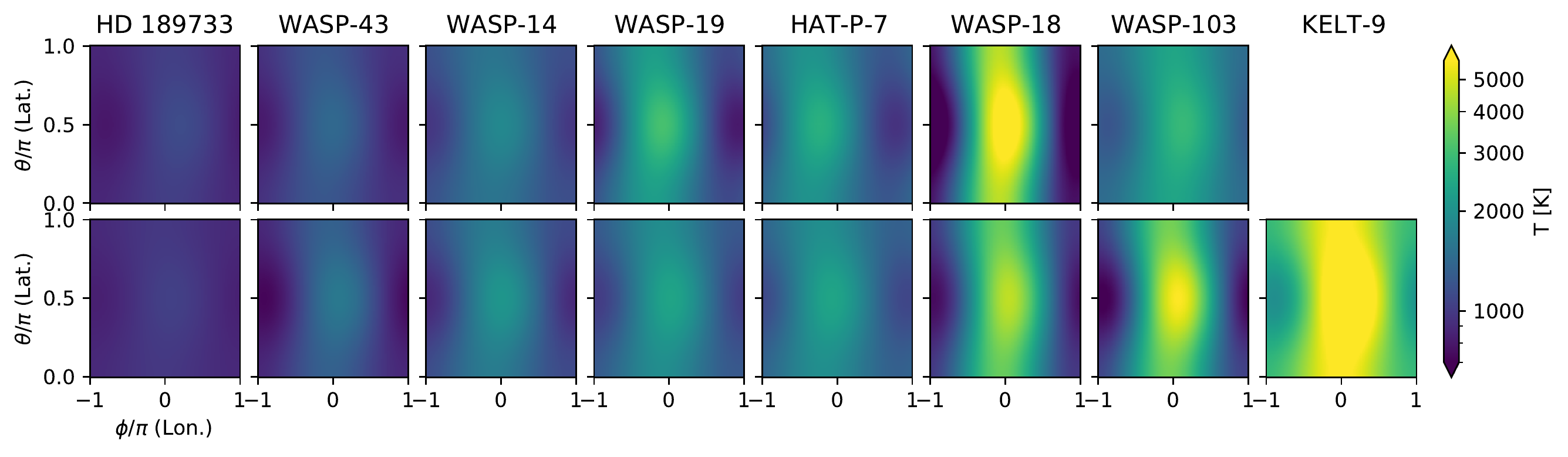}
    \caption{Maximum-likelihood temperature maps for several of the hottest exoplanets, in order of increasing equilibrium temperature from left to right, where the horizontal axes are planetary longitude and vertical axes are latitude. The horizontal axis is the longitude centered on the substellar point, the vertical axis is latitude from pole to pole. The upper row shows the \spitzer IRAC Channel 1 (3.6 $\rm \mu$m) inferences, the lower row shows IRAC Channel 2 (4.5 $\rm \mu$m). The longitudinal temperature variations are constrained by the \spitzer phase curve observations while the latitudinal temperature variations are fixed at the parameterization that best describes the GCMs.}
    \label{fig:maps}
\end{figure*}

\begin{figure}
    \centering
    \includegraphics[width=\columnwidth]{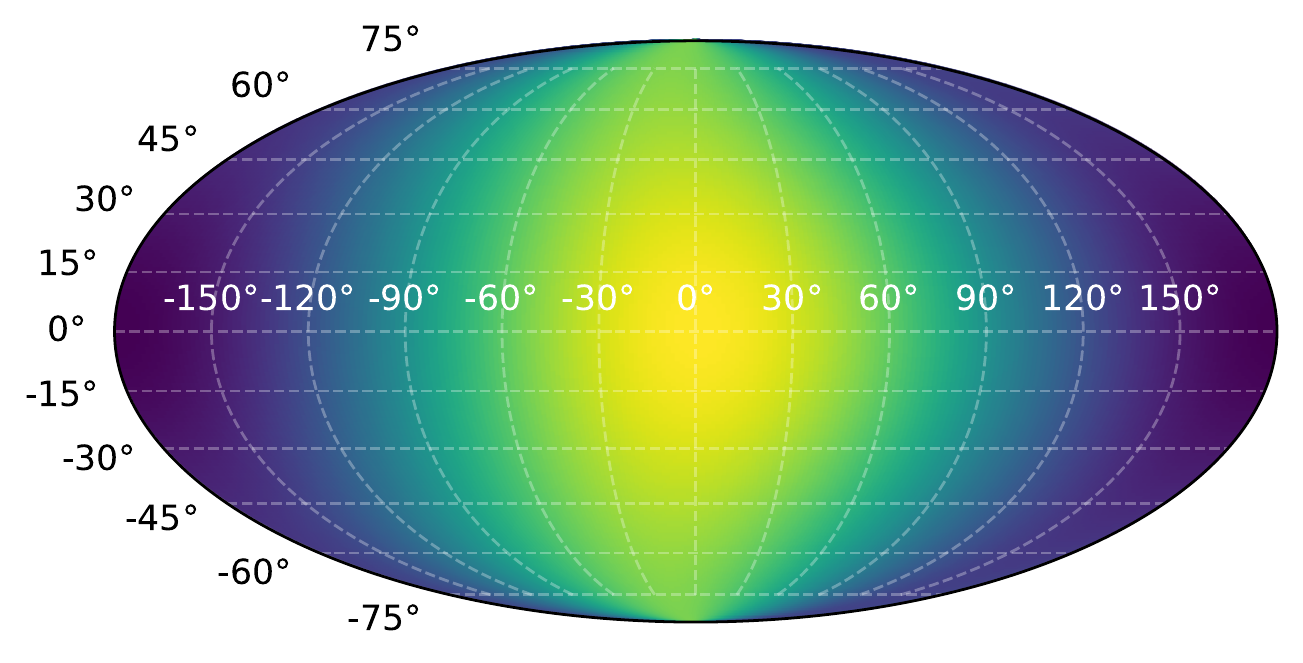}
    \caption{The ``universal'' temperature map for hot Jupiters: scaled versions of this temperature map are consistent with both the temperature distributions in GCMs (which constrain the latitudinal distribution) and the phase curves of the hottest exoplanets (which constrain the longitudinal distribution). This map has $\alpha=0.6$ and $\omega_\mathrm{drag} = 4.5$ and an arbitrary scaling for $f$, $C_{11}$ and $\Delta\phi=0$. }
    \label{fig:map_universal}
\end{figure}

\section{Discussion} \label{sec:discussion}

\subsection{Comparison with \citet{Bell2021}}

\begin{figure}
    \centering
    \includegraphics[width=\columnwidth]{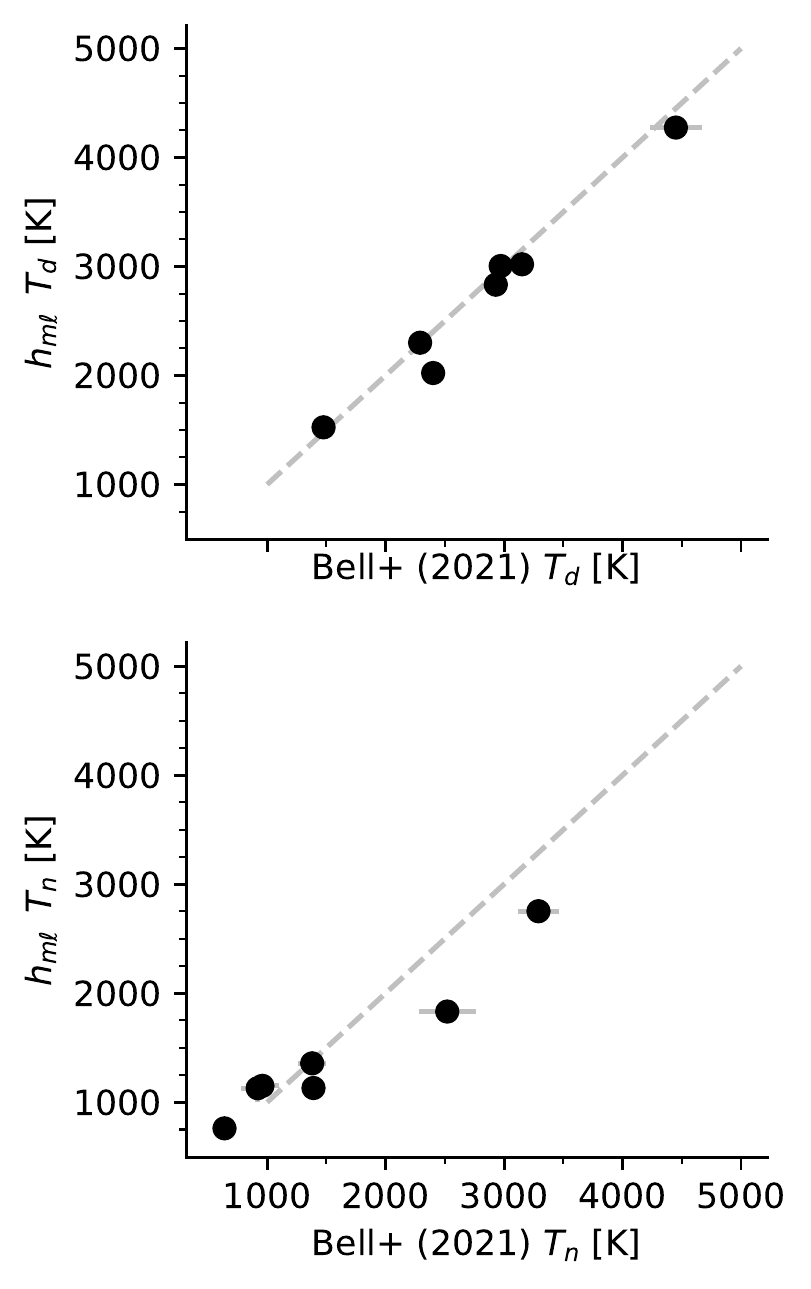}
    \caption{Hemispherically-averaged dayside and nightside temperatures from analysis with the \hml framework compared with a more traditional analysis by \citet{Bell2021}, showing good agreement between techniques.}
    \label{fig:temp_compare}
\end{figure}

Recently \citet{Bell2021} produced re-reduced \spitzer phase curves of many targets, including seven planets in common with this analysis. We compare their results with the ones presented in this work in Figure~\ref{fig:temp_compare}, showing excellent agreement on the dayside temperatures, indicating that the dayside temperature inferences are not strongly model dependent. The two hottest planets show significantly hotter temperatures in the \citet{Bell2021} analysis than in the analysis presented here, which may be attributed to differences in the reduction process.

\subsection{Temperature forecasts for hot Jupiters}

By tightly constraining the $\alpha$ and $\omega_\mathrm{drag}$ parameters we have essentially fit the phase curves of each planet with a nearly identical temperature map with different relative scalings and rotations. Figure \ref{fig:best_params} shows roughly monotonic trends for the retrieved values of $\Delta \phi$ and $C_{11}$ as a function of the equilibrium temperature for our analysis of both GCMs and Spitzer data.  These trends suggest a well-behaved sequence of \spitzer-derived temperature maps, which we show in Figure \ref{fig:maps}.  This sequence in turn suggests, to first order, a roughly universal climate state for hot Jupiters, consistent with theoretical expectations \citep{Showman2011, Tsai2014}. 

In other words, {\it one} unified map can explain the phase variations of all of the hot Jupiters, if we rescaled all of the derived maps by the three parameters ($f$, $\Delta \phi$ and $C_{11}$). We propose that most hot Jupiters should have temperature maps like Figure~\ref{fig:map_universal}. We can test this hypothesis with upcoming observations from the CHEOPS space telescope, which will measure a precise and time-resolved phase curve of KELT-9 b in the optical, among other planets. Deviations from this ``mean hot Jupiter'' temperature map could be indicative of variations in the atmospheric structure with depth, and time-varying atmospheric conditions, for example.

\subsection{When to use \hml to fit a phase curve}

The \hml basis is a generalized spherical harmonic expansion representation (parabolic cylinder functions) of the temperature map of a celestial body based on solutions to the shallow water system. In this work, we have shown that spherical harmonic degrees as small as $\ell_\mathrm{max} = 1$ are sufficient to approximate the temperature maps of highly irradiated, giant exoplanets, for certain values of $\alpha$ and $\omega_\mathrm{drag}$. 

The shallow water framework is applicable for systems where the horizontal dimension of dynamics in the fluid far exceeds the radial one (hence the ``shallow'' part of the name). This is true for giant exoplanets, for example, where a few pressure scale heights in the outer parts of the atmosphere interact with light, and the bulk of the planetary radius is insulated from stellar irradiation. 

The shallow water system at small spherical harmonic degrees would struggle to approximate the temperature maps of planets without atmospheres. Such worlds can have $T_d \gg 0$ K and $T_n \approx 0$ K when synchronously rotating, and the discontinuity near the terminator could only be successfully reproduced by the \hml basis at large spherical harmonic degrees. We note that this challenge is not unique to the \hml basis, but rather a shortcoming of spherical harmonic expansion maps in general. As such, we recommend that the \hml basis should be applied only to planets that are likely to have atmospheres.

\subsection{Discrepancies between bandpasses} \label{sec:discussion-bandpass}

Differences between the inferred parameters across multiple filter bandpasses can be attributed to variations in the typical altitude probed by the photosphere of the planet at each wavelength. Table~\ref{tab:spitzer} shows the inferred phase curve parameters using the \hml basis for \spitzer IRAC Channels 1 and 2 for eight planets. The dayside and nightside temperatures, for example, can be significantly different between the two \spitzer bandpasses. This is the case for WASP-18 b, which has a dayside integrated temperature nearly 250 K cooler in Channel 1 than Channel 2, corresponding to a difference of $\approx10\times$ the uncertainty on either measurement. Such differences between bandpasses can occur when variable opacity with wavelength determines the depth of the photosphere. The 3D nature of the planet's atmosphere is often described with a piecewise, non-linear temperature--pressure profile with altitude \citep[see, e.g.:][]{Kitzmann2020}, and the phase curve brightness temperatures presented here represent samples from the T--P profile weighted by the contribution function for a given bandpass.

\subsection{Reporting Bond albedos and heat redistribution from 2D temperature maps}

As discussed in Section~\ref{sec:albedos}, \citet{Cowan2011b} devised a simple parameterization for the Bond albedo $A_B$ and heat redistribution efficiency parameter $\varepsilon$. This ``0D'' approach constructs a planet with a dayside and nightside temperature, and the relationship between those temperatures is dictated by the energy balance in the star-planet system, which is parameterized by $A_B$ and $\varepsilon$. We show in  Section~\ref{sec:albedos} that the \citet{Cowan2011b} relations for these parameters can be manipulated to compute $A_B$ and $\varepsilon$ given the temperature map of the planet. 

Since the temperature maps inferred in this work are 2D, we outline a more general relationship between the full temperature map, $A_B$, and $\varepsilon$ in Section~\ref{sec:general-A_B}. This 2D, energy-balanced approach sets the thermal emission of a planet through its entire surface equal to the radiation incident on the planet's dayside. We adopt the resulting expression for $A_B$ throughout this work. 

\begin{figure}
    \centering
    \includegraphics[width=\columnwidth]{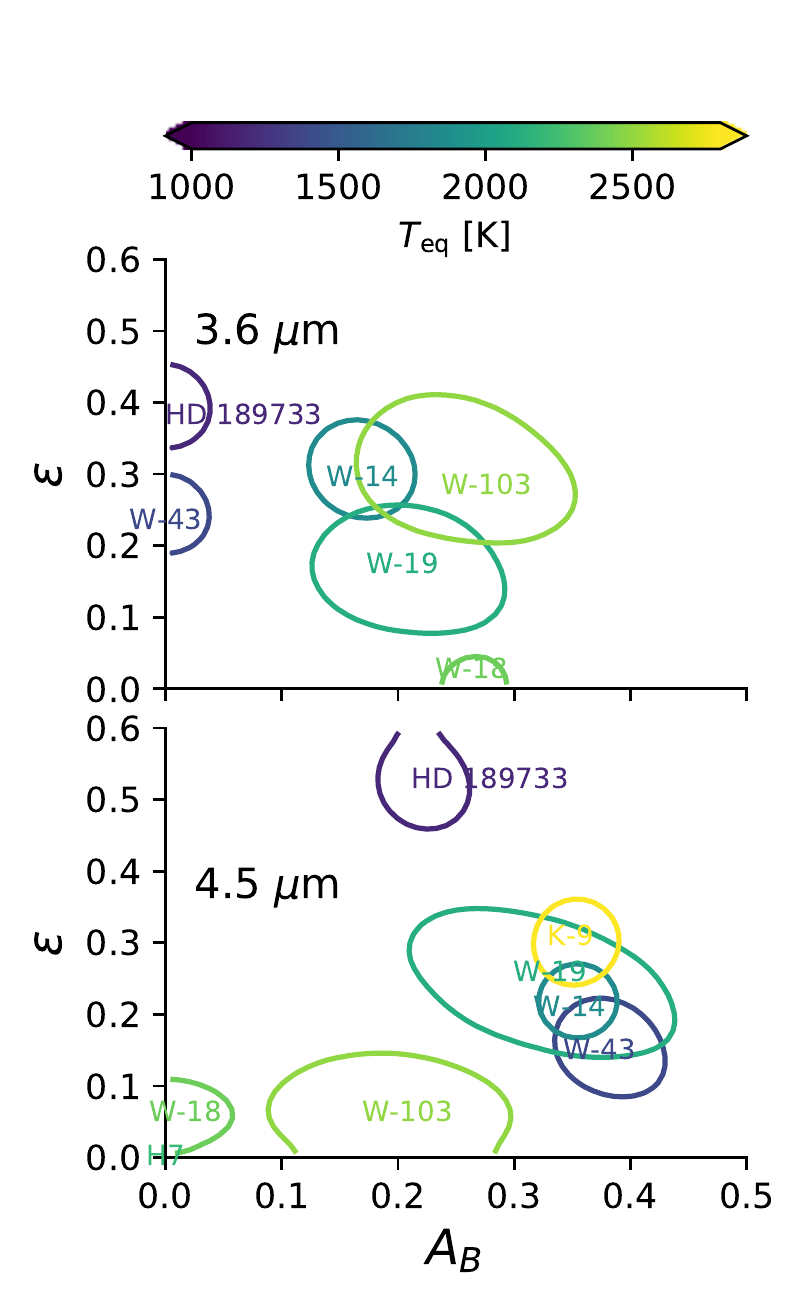}
    \caption{Posterior distributions for the Bond albedos and heat redistribution efficiencies from the \spitzer phase curves of several hot exoplanets, colored by their equilibrium temperatures, in the style of Figure 3 of \citet{Schwartz2017}. Section~\ref{sec:albedos} defines $A_B$ and $\varepsilon$ for the 2D temperature maps inferred from the \hml basis.}
    \label{fig:ab_eps}
\end{figure}

We also redefine the heat redistribution efficiency as the ratio of the flux emitted by the nightside of the planet and the flux emitted through the dayside. This definition preserves the limits $\varepsilon \in [0, 1]$, as in \citet{Cowan2011b}, and is straightforward to compute from both GCM outputs and \hml maps. The Bond albedos and heat redistribution parameters derived in this work are plotted in Figure~\ref{fig:ab_eps}.

\subsubsection{On the definition of the Bond albedo}

The Bond albedo is defined as the spherical albedo $A_s(\lambda)$ integrated over all wavelengths, weighted by the stellar spectrum \citep{Marley1999},
\begin{equation}
    A_B = \frac{\int I_\star ~A_s~d\lambda}{\int I_\star~d\lambda}, \label{eqn:AB}
\end{equation}
where $I_\star(\lambda,T_\star)$ is the incoming stellar radiation. In Section~\ref{sec:general-A_B}, we express the Bond albedo assuming the brightness temperature of the planet in any finite bandpass will balance the incoming stellar radiation over all wavelengths; in other words, the planet is assumed to be a blackbody\footnote{We note that this definition of a ``blackbody'' is somewhat inconsistent, since a blackbody should be perfectly absorbing with $A_B = 0$.} which reflects light with some efficiency $1 - A_B$. A real planet does not emit as a blackbody and has spectral features, and thus its brightness temperatures in finite bandpasses are not equivalent to the equilibrium temperature. As a result of these subtleties, (1) the $A_B$ inferred from a finite bandpass may be negative, and (2) $A_B$ estimates inferred from brightness temperatures in distinct bandpasses may have different values\footnote{In this sense, the ``$A_B$'' we discuss in this manuscript {\it is} a function of wavelength, unlike its definition in Equation~\ref{eqn:AB}.}.

\subsubsection{On the definition of the heat redistribution efficiency}

The heat redistribution efficiency $\varepsilon$ as defined for 2D temperature maps in Equation~\ref{eqn:varepsilon2d} is distinct from the definition in \citet{Cowan2011b}. For example, consider a planet that has a uniform dayside temperature $T_d$ and uniform nightside temperature $T_n$. The 0D derivation yields
\begin{equation}
    \varepsilon_{\rm 0D} = \frac{8 T_n^4}{3T_d^4 + 5T_n^4}, \label{eqn:eps0d}
\end{equation}
whereas the 2D approach reduces to
\begin{equation}
    \varepsilon_{\rm 2D} = \frac{T_n^4}{T_d^4}.
\end{equation}
These expressions are equivalent when $\varepsilon_{\rm 0D} = \varepsilon_{\rm 2D} = 1$ and $T_d = T_n$. For imperfect recirculation efficiencies, the appearance of the dayside and nightside temperatures in the denominator of Equation~\ref{eqn:eps0d} ensures that the 0D and 2D approaches yield different efficiencies. It can be shown that the 0D efficiencies are larger than the 2D efficiencies by up to $\lesssim 0.24$, where the maximal difference occurs for $T_d+T_n = 8(T_d - T_n)$.

\subsection{Reflected light} \label{sec:reflected}

\begin{figure*}
    \centering
    \includegraphics[width=\textwidth]{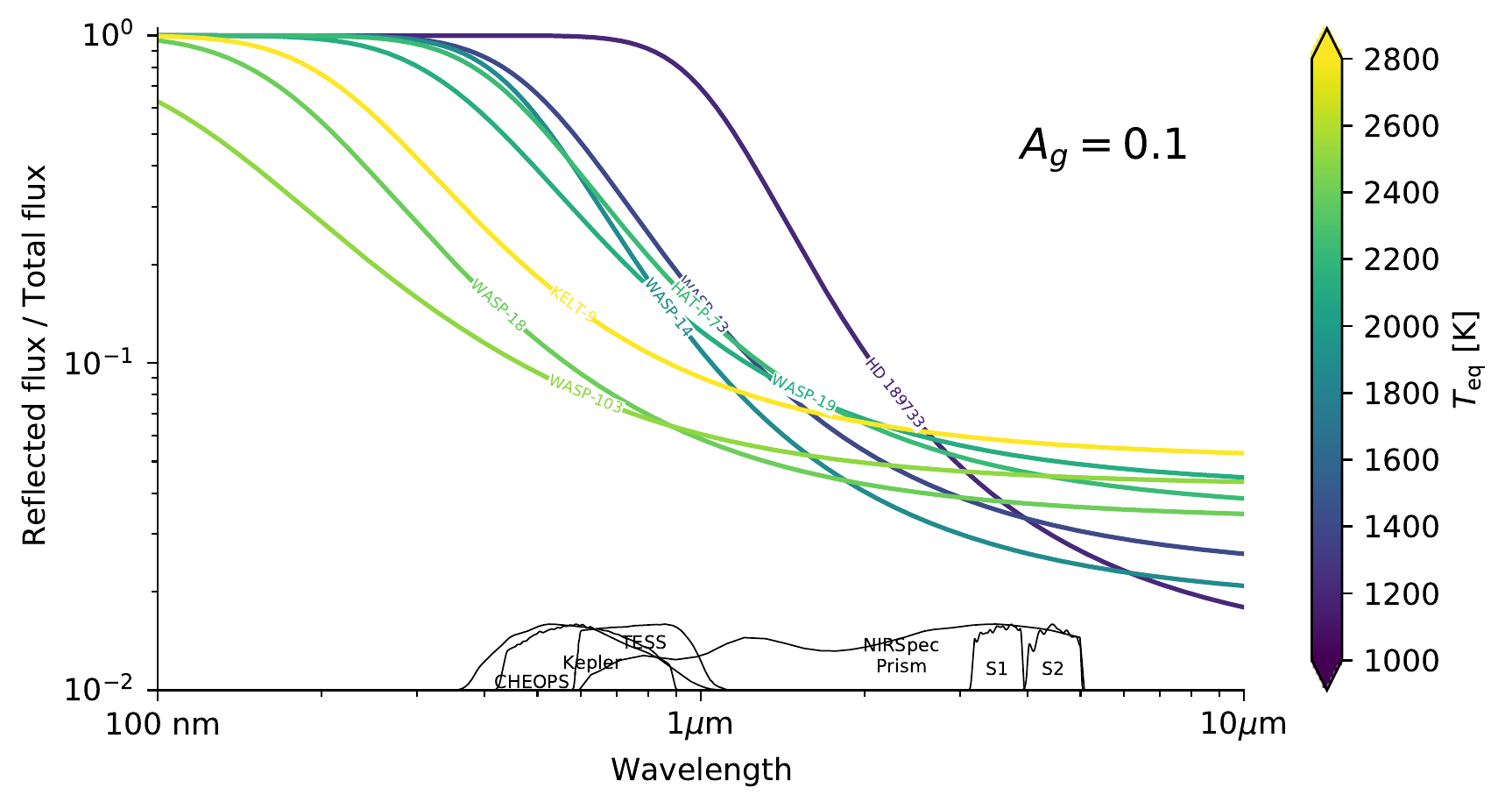}
    \caption{Fraction of light reflected from the planetary surface at secondary eclipse normalized by the total flux from both reflection and thermal emission, assuming constant $A_g (\lambda)= 0.1$, after Figure 4 of \citet{Schwartz2015}. Blackbody emission is assumed for both the planet and the star. See Appendix~\ref{app:c} for details.}
    \label{fig:reflected}
\end{figure*}

One caveat of the analysis presented here is that we make the usual assumption that the reflected light component of the phase curve in the \spitzer bandpass is negligible. Figure~\ref{fig:reflected} shows the expected contribution of reflected light to the observed, reflected and thermal emission spectrum of the planet, assuming $A_g=0.1$, similar to typical hot Jupiters in the \kepler bandpass \citep{Heng2013}. Details of the calculation are provided in Appendix~\ref{app:c}.

The contribution of reflected light to the total planetary flux is on the order of a few percent for the warm \spitzer bandpasses for all targets in this work. Reflected light contributes significant flux at optical wavelengths for cooler exoplanets such as HD 189733 b and WASP-43 b. Joint inferences are required for the thermal emission and reflectance properties of cooler exoplanets at optical wavelengths \citep[see, e.g.:][]{Heng2021}. Ultra-hot Jupiters such as KELT-9 b are dominated by thermal emission even at blue-optical wavelengths. Phase curves of cooler giant exoplanets from the James Webb Space Telescope (JWST) in the NIRSpec/Prism mode, for example, will be a mixture of thermal emission and reflected light that varies with wavelength. A similar conclusion on reflected light was reached by \citet{Keating2017}.

\subsection{Comparison with traditional spherical harmonics}

 include in Appendix~\ref{app:d} calculations which indicate that the \hml basis reproduces GCM temperature maps more accurately than the traditional spherical harmonic \ylm basis. We conclude that the \hml basis functions provide better fits to GCM temperature maps than similar \ylm basis functions, which may be relevant when self-consistently computing hotspot offsets and phase curve amplitudes directly from GCM temperature maps and indirectly from phase curves, as is the subject of this work.

\section{Conclusion} \label{sec:conclusion}

\begin{itemize}
    \item GCM temperature maps and real photometric phase curves can be compared using the same language, which is a set of modified spherical harmonic basis functions which we call the \hml basis (Section~\ref{sec:model})
    \item The ``0D'' derivation for the Bond albedo $A_B$ and the heat redistribution efficiency $\varepsilon$ popularized by \citet{Cowan2011b} can be generalized when considering 2D temperature maps (Section~\ref{sec:albedos})
    \item There are only three free thermal parameters required to fit each phase curve in a given bandpass, $\{C_{11}, \Delta\phi, f\}$
    \item We can fit GCM outputs (Section~\ref{sec:gcms}) and \spitzer phase curves (Section~\ref{sec:spitzer})  with the same $\{\alpha, \omega_\mathrm{drag}\}$, and therefore there exists one ``universal'' hot Jupiter map, which after scaling and rotation can matches existing hot Jupiter observations
    \item Bond albedos, heat redistribution efficiencies, phase offsets, and day- and night-side temperatures are computed for several hot Jupiters with \spitzer phase curves (Table~\ref{tab:spitzer})
    \item We confirm that reflected light from planets will contaminate the bluest wavelengths accessible to the James Webb Space Telescope, and the contamination is most severe for cooler planets
    \item Comparisons of the \hml basis with the traditional \ylm basis suggest that GCMs are more accurately reproduced with the \hml basis, though the resulting phase curves are usually indistinguishable.
    \item We offer code that computes the \hml maps and their phase curves efficiently called \textsf{kelp}, which is publicly
    available with implementations in \textsf{Cython}, \textsf{theano} and \textsf{JAX}\footnote{\url{https://github.com/bmorris3/kelp}}
\end{itemize}

\begin{acknowledgements}
This manuscript was carefully and thoughtfully reviewed and improved by Nicolas Cowan. This work has been carried out in the framework of the PlanetS National Centre of Competence in Research (NCCR) supported by the Swiss National Science Foundation (SNSF). We gratefully acknowledge software testing and discussions with Matthew Hooton and Adrien Deline. KH acknowledges partial financial support by a European Research Council Consolidator Grant (number 771620, project EXOKLEIN). KJ acknowledges financial support by a Hans Sigrist Foundation Ph.D Fellowship.  We thank Russell Deitrick, Elsie Lee and Nathan Mayne for providing GCM output from previous publications. This research has made use of NASA's Astrophysics Data System. This paper includes data collected by the TESS mission. Funding for the TESS mission is provided by the NASA Explorer Program. This research has made use of the VizieR catalogue access tool, CDS, Strasbourg, France (DOI: 10.26093/cds/vizier). The original description of the VizieR service was published in A\&AS 143, 23. We gratefully acknowledge the open source software that made this work possible: \textsf{astropy} \citep{Astropy2013, Astropy2018}, \textsf{ipython} \citep{ipython}, \textsf{numpy} \citep{numpy}, \textsf{scipy} \citep{scipy}, \textsf{matplotlib} \citep{matplotlib}, \textsf{batman} \citep{Kreidberg2015}, \textsf{emcee} \citep{Foreman-Mackey2013}, \textsf{PyMC3} \citep{pymc3}, \textsf{theano} \citep{theano}, \textsf{pymc3-ext} \citep{Foreman-Mackey2021}, \textsf{corner} \citep{Foreman-Mackey2016}, \textsf{cython} \citep{behnel2011cython}, \textsf{JAX} \citep{jax2018github}.
\end{acknowledgements}

\bibliographystyle{aa} 
\bibliography{bibliography} 

\appendix

\section{\hml linear solve} \label{app:a}

In this Appendix we outline how to reproduce a given GCM temperature map, for example, with the \hml basis, as done in the bottom row of Figure~\ref{fig:hottest_maps}. We outline here the linear solution for the $C_{m\ell}$ terms that produce \hml maps at arbitrary \lmax, following the linear solve technique of \citet{Luger2019}. 

For all $m,\ell$, one can construct the \hml perturbation map from Equation~\ref{eqn:hml}, setting each $C_{m\ell}$ to a positive constant. Each 2D perturbation map is then flattened into a column vector, and each column vector is stacked with the others to produce a design matrix; this is known in the \textsf{starry} framework as a ``pixelization matrix,'' ${\bf P}$. We stack onto this design matrix a column vector of ones to represent the mean temperature. One can then linearly solve for ${\bf Q}$ in
\begin{equation}
{\bf A} {\bf Q} = {\bf P}^\top {\bf w}^2,
\end{equation}
where ${\bf w}$ is a weight matrix which can be assigned to $\sin{\theta}$ in order to reproduce equatorial features and down-weight the polar regions, for example, and the matrix ${\bf A}$ is
\begin{equation}
    {\bf A} = {\bf P}^\top {\bf w}^2 {\bf P} + \epsilon {\bf I},
\end{equation}
where ${\bf I}$ is the identity matrix and $\epsilon$ is a small positive constant. The vector of unnormalized $C_{m\ell}$ coefficients $c^\prime$ which reproduce a given temperature map ${\bf T}_\mathrm{input}$ can be found by multiplying ${\bf Q}$ by the temperature map,
\begin{equation}
    c^\prime = {\bf Q T}_\mathrm{input}.
\end{equation}
Finally we normalize each $c^\prime$ by the mean of the temperature map $c = c^\prime / \bar{\bf T}_\mathrm{input}$. The resulting $c$ vector represents each $C_{m,\ell}$ power coefficient to be passed through the \hml framework in order to produce the \hml representation of ${\bf T}_\mathrm{input}$.

\section{Reflected light} \label{app:c}

The observed flux from an exoplanet is the sum of the thermal emission spectrum and the reflected starlight, 
\begin{equation}
    F_{p}(\lambda, \xi) = F_\mathrm{refl}(\lambda, \xi) + F_\mathrm{therm}(\lambda, \xi).
\end{equation}
The thermal emission component is written in Equation~\ref{eqn:diskint}.

The reflected light component is defined as a function of the geometric albedo at a given wavelength
\begin{equation}
    F_\mathrm{refl}(\lambda, \xi) = A_{g}(\lambda) \left(\frac{R_p}{a}\right)^2 \Psi(\xi) F_\star(\lambda),
\end{equation}
where $\Psi(\xi)$ is the integral phase function and $F_\star(\lambda)$ is the stellar emission spectrum. 

The ratio of the reflected light spectrum observed at secondary eclipse ($\xi=0$) to the total planetary flux is given by
\begin{eqnarray}
    \frac{F_\mathrm{refl}}{F_p}(\lambda, 0) =& \left(1 + \frac{F_\mathrm{therm}(\lambda, 0)}{A_{g}(\lambda) \left(\frac{R_p}{a}\right)^2 F_\star(\lambda),} \right)^{-1},\\
    =& \left(1 + \left(\frac{a}{R_\star}\right)^2\frac{\iint  \mathcal{B}_\lambda(T_p) \cos(\phi) \sin^2(\theta) d\phi d\theta }{\pi A_{g}(\lambda) \mathcal{B}_\lambda(T_\star)} \right)^{-1}.
\end{eqnarray}

In Figure~\ref{fig:reflected}, we compute each curve by solving the equation above with the corresponding planetary and stellar parameters, assuming $A_g(\lambda) = 0.1$ is constant with wavelength. Note that the thermal emission component is computed using the temperature map inferred from the \spitzer Channel 2 phase curves, though as discussed in Section~\ref{sec:discussion-bandpass}, the observed, brightness temperature of the planet is a function of wavelength. 

\section{Comparison with traditional spherical harmonics} \label{app:d}

\begin{figure*}
    \centering
    \includegraphics[width=\textwidth]{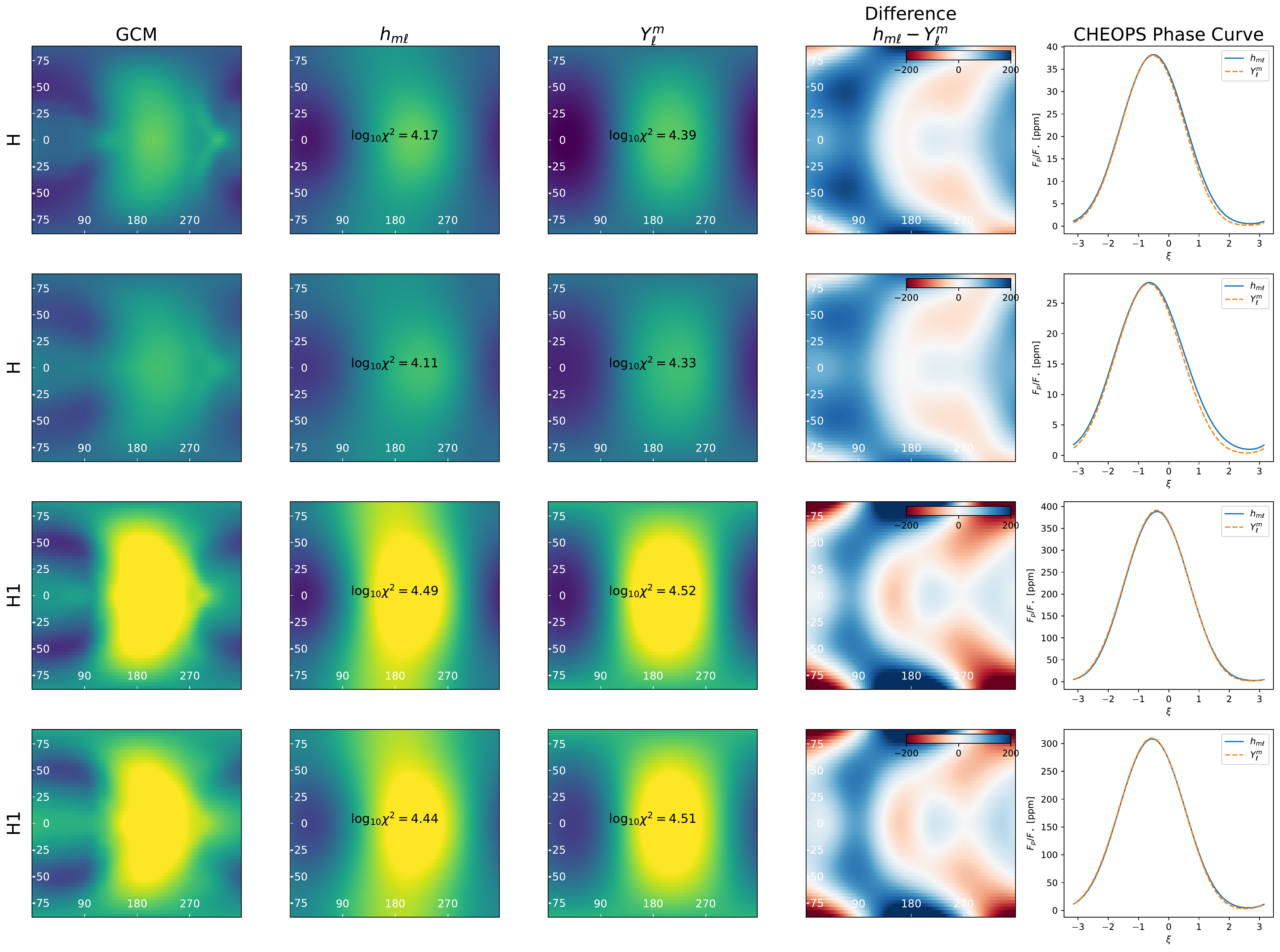}
    \caption{The four hottest GCM temperature maps (first column), the best-fit linear solutions with a low-order \hml representation (second column), the fits with low-order ordinary spherical harmonic representation (third column, labeled \ylm), the difference between the two representations (fourth column), and the phase curves produced by the two representations of the GCM map as observed in the CHEOPS bandpass (fifth column).}
    \label{fig:ylm}
\end{figure*}

In Figure~\ref{fig:ylm}, we select a model with three free parameters in both the \hml and \ylm representations to solve linearly (see our Appendix~\ref{app:a} for the mathematical approach). The differences between \hml and \ylm can be as large as 400 K peak-to-peak. The $\chi^2$ between both the \hml and \ylm representations compared with the GCM temperature map is labelled on each \hml and \ylm map. We note that the $\chi^2$ is significantly smaller for each \hml representation than the corresponding \ylm representation with the same number of free parameters. This indicates that the GCM temperature maps are more accurately reproduced by the \hml basis than the \ylm basis, even though the phase curves produced by either the \hml or \ylm solutions are nearly indistinguishable.

\end{document}